\PassOptionsToPackage{pdfpagelabels=false}{hyperref} 
\RequirePackage{silence} 

\documentclass[fleqn,usenatbib]{mnras}

\usepackage{newtxtext,newtxmath}
\usepackage[T1]{fontenc}
\usepackage{ae,aecompl}
\usepackage{graphicx}
\usepackage{amsmath}
\usepackage{array}
\usepackage[dvipsnames]{xcolor}
\usepackage{xspace}

\definecolor{olive}{rgb}{0, 0.7, 0}

\newcommand{\delight}{\textsc{delight}\xspace}
\newcommand{\bcnz}{\textsc{bcnz}$2$\xspace} 
\newcommand{\bpz}{\textsc{bpz}\xspace}
\newcommand{\lephare}{\textsc{lephare}\xspace}

\newcommand{\annztwo}{\textsc{annz}$2$\xspace}
\newcommand{\deepz}{\textsc{deepz}\xspace}

\newcommand{\photoz}{photo-$z$\xspace}
\newcommand{\Photoz}{Photo-$z$\xspace}
\newcommand{\specz}{spec-$z$\xspace}

\newcommand{\rf}[1]{\textrm{#1}}
\newcommand{\tf}[1]{\texttt{#1}}

\title[PAUS: narrowband photo-z's using GPs]{The PAU Survey: narrowband photometric redshifts using Gaussian processes}

\author[J. Y. H. Soo et al.]{
John Y. H. Soo$^{1,2}$\thanks{E-mail: johnsooyh@usm.my}, Benjamin Joachimi$^{2}$, Martin Eriksen$^{3}$, Ma\l gorzata Siudek$^{3,4}$, \and Alex Alarcon$^{5}$, Laura Cabayol$^{3}$, Jorge Carretero$^{3,6}$, Ricard Casas$^{7,8}$, \and Francisco J. Castander$^{7,8}$, Enrique Fern\'andez$^{3}$, Juan Garc\'ia-Bellido$^{9}$, \and Enrique Gaztanaga$^{7,8}$, Hendrik Hildebrandt$^{10}$, Henk Hoekstra$^{11}$, Ramon Miquel$^{3,12}$, \and Cristobal Padilla$^{3}$, Eusebio S\'anchez$^{13}$, Santiago Serrano$^{7,8}$ and Pau Tallada-Cresp\'i$^{6,13}$.
\\
$^{1}$School of Physics, Universiti Sains Malaysia (USM), 11800 USM, Pulau Pinang, Malaysia.\\
$^{2}$Department of Physics and Astronomy, University College London (UCL), Gower Street, London WC1E 6BT, UK.\\
$^{3}$Institut de F\'{i}sica d'Altes Energies (IFAE), The Barcelona Institute of Science and Technology, 08193 Bellaterra (Barcelona), Spain.\\
$^{4}$National Centre for Nuclear Research, 7 Pasteura str., 02-093 Warsaw, Poland.\\
$^{5}$High Energy Physics Division, Argonne National Laboratory, Lemont, IL 60439, USA.\\
$^{6}$Port d'Informaci\'{o} Cient\'{i}fica (PIC), Universitat Aut\`onoma de Barcelona, Carrer Albareda S/N, 08193 Bellaterra (Barcelona), Spain.\\
$^{7}$Institute of Space Sciences (ICE/CSIC), Universitat Aut\`onoma de Barcelona, Carrer de Can Magrans S/N, 08193 Cerdanyola del Vall\`es (Barcelona), Spain.\\
$^{8}$Institut d'Estudis Espacials de Catalunya (IEEC), 08034 Barcelona, Spain.\\
$^{9}$Instituto de F\'isica Te\'orica (IFT-UAM/CSIC), Universidad Aut\'onoma de Madrid, 28049 Madrid, Spain.\\
$^{10}$German Centre for Cosmological Lensing, Astronomisches Institut, Ruhr-Universit\"at Bochum (AIRUB), Universit\"atsstr. 150, 44801 Bochum, Germany.\\
$^{11}$Leiden Observatory, Leiden University, Niels Bohrweg 2, 2333 CA, Leiden, The Netherlands.\\
$^{12}$Instituci\'o Catalana de Recerca i Estudis Avan\c{c}ats (ICREA), 08010 Barcelona, Spain.\\
$^{13}$Centro de Investigaciones Energeticas, Medioambientales y Tecnologicas (CIEMAT), Avenida Complutense 40, 28040 Madrid, Spain.\\
}

\date{Accepted XXX. Received YYY; in original form ZZZ}

\pubyear{2020}

\begin{document}
\label{firstpage}
\pagerange{\pageref{firstpage}--\pageref{lastpage}}
\maketitle

\begin{abstract}
We study the performance of the hybrid template-machine-learning photometric redshift (\photoz) algorithm \delight, which uses Gaussian processes, on a subset of the early data release of the Physics of the Accelerating Universe Survey (PAUS). We calibrate the fluxes of the $40$ PAUS narrow bands with $6$ broadband fluxes ($uBVriz$) in the COSMOS field using three different methods, including a new method which utilises the correlation between the apparent size and overall flux of the galaxy. We use a rich set of empirically derived galaxy spectral templates as guides to train the Gaussian process, and we show that our results are competitive with other standard photometric redshift algorithms. \delight achieves a \photoz $68$th percentile error of $\sigma_{68}=0.0081(1+z)$ without any quality cut for galaxies with $i_\rf{auto}<22.5$ as compared to $0.0089(1+z)$ and $0.0202(1+z)$ for the \bpz and \annztwo codes, respectively. \delight is also shown to produce more accurate probability distribution functions for individual redshift estimates than \bpz and \annztwo. Common \photoz outliers of \delight and \bcnz (previously applied to PAUS) are found to be primarily caused by outliers in the narrowband fluxes, with a small number of cases potentially indicating spectroscopic redshift failures in the reference sample. In the process, we introduce performance metrics derived from the results of \bcnz and \delight, allowing us to achieve a \photoz quality of $\sigma_{68}<0.0035(1+z)$ at a magnitude of $i_\rf{auto}<22.5$ while keeping $50$ per cent objects of the galaxy sample.
\end{abstract}


\section{Introduction} \label{sec:intro}
Photometric redshift (\photoz) estimation continues to be an active research area as it plays a major role in solving the big questions in cosmology. Redshifts provide radial information (distance) to the traditional two dimensional sky maps of galaxies. They are traditionally determined through spectroscopic methods (spectroscopic redshifts, or \specz's). Yet since the process requires long telescope time for high completeness, \photoz's are instrumental for the analysis of large surveys containing of order $10^{8-9}$ galaxies. \Photoz methodology has been evolving and improving a lot over the past couple of decades \mbox{\citep[e.g.][]{brescia_data_2018,salvato_many_2019}}, such that it had been sufficiently useful for most recent cosmological researches. 

\Photoz, as its name suggests, is often determined through the use of a handful of broadband photometric filters obtained from large sky surveys. \Photoz estimation methods are generally categorised into two different types: the template-based method, which relies on accurate models of spectral energy distribution (SED) templates of different types of galaxies; and the data-driven empirical method, which relies on training sets of galaxies and machine-learning algorithms. Each method however has its own limitations: template-based methods may produce \photoz's with large scatter and catastrophic rates without representative templates; while machine-learning methods may perform poorly outside the regions of the parameters covered by the training sample \citep{disanto_return_2018}. As a result, hybrid methods have been implemented to utilise the best of both worlds \citep{cavuoti_cooperative_2017,duncan_photometric_2018,duncan_observational_2019}.

Many current and upcoming surveys such as the Dark Energy Survey \citep[DES,][]{abbott_dark_2005}, Legacy Survey of Space and Time \citep[LSST,][]{ivezic_lsst:_2008}, \textit{Euclid} \citep{laureijs_euclid_2011}, Kilo-Degree Survey \citep[KiDS,][]{de_jong_kilo-degree_2013}, Wide Field Infrared Survey Telescope \citep[WFIRST,][]{spergel_wide-field_2013} and Hyper Suprime-Cam \citep[HSC,][]{aihara_hyper_2018} have set stringent \photoz requirements to ensure that they meet their science goals, forcing the quality of \photoz methodology to constantly improve. For example, LSST's \photoz requirement is to reach a root-mean-square error of $\sigma_\rf{RMS}<0.02(1+z)$, while the \textit{Euclid} requirement is $\sigma_\rf{RMS}<0.05(1+z)$. High quality \photoz's are required for a reliable estimation of e.g. weak lensing \citep{benjamin_cfhtlens_2013}, angular clustering \citep{crocce_galaxy_2016}, intrinsic alignment \citep{johnston_pau_2020}, structure formation, galaxy classification and galaxy properties \citep{jouvel_photometric_2017,laigle_cosmos2015_2018,siudek_vimos_2018_1}.

The aforementioned surveys are predominantly broadband surveys which use between $4$-$9$ broadband filters ranging from infrared to ultraviolet. This work however, explores the estimation of \photoz's in narrowband surveys, focusing on the Physics of the Accelerating Universe Survey \citep[PAUS,][]{padilla_physics_2019}, which observes the sky using $40$ narrow bands (see Section~\ref{sec:pau}). Producing high quality \photoz's for such a survey requires careful optimisation between narrow and broad bands, since machine-learning based methods have to be optimised for a larger number of inputs \citep{eriksen_pau_2020}, while template-based methods require more attention towards the narrow emission line features. 

\citet{marti_precise_2014} used simulations to predict that by using PAUS narrowband photometry, the \photoz quality could reach an unprecedentedly low $68$th percentile error of $\sigma_{68}=0.0035(1+z)$ at a quality cut of $50$ per cent at $i<22.5$. This has been verified by \citet{eriksen_pau_2019}, where they combined the $40$ PAUS narrow bands (early data release) with broad bands $uBVriz$ from the Cosmic Evolution Survey \citep[COSMOS,][]{laigle_cosmos2015_2016}, and using their template-based \photoz code \bcnz, they showed that this result is achievable when a $50$ per cent photometric quality cut was imposed on the final testing set. In a more recent work, \citet{eriksen_pau_2020} used \textsc{deepz}, a deep learning algorithm on the same data set and showed that it outperformed \bcnz by reaching $50$ per cent lower in $\sigma_{68}$. Furthermore, \citet{alarcon_pau_2021} showed that an ever greater precision can be achieved when using additional photometric bands available in the COSMOS field (a total of $66$ bands).

We are motivated by the work of \citet{eriksen_pau_2019}, but instead of using purely template-based methods, we attempt to achieve this PAUS \photoz precision by utilising Gaussian processes (GPs, see Section~\ref{sec:delight}) to make empirical adjustments to templates, working on the same data set and conditions. We seek to produce an independent method that is competitive, as that will allow us to exploit synergies with \bcnz by \citet{eriksen_pau_2019} as shown in this work, \textsc{deepz} \citep{eriksen_pau_2020}, and \photoz's by \citet{alarcon_pau_2021} in the future. Therefore the contents of this paper reflect our findings, putting special emphasis on the performance and application of \delight \citep{leistedt_data-driven_2017}, a hybrid template-machine-learning \photoz code. When carefully calibrated and combined with COSMOS broadband fluxes, \delight should achieve equally good results as that of \bcnz. The main aims of this paper are threefold:

\begin{enumerate}
    \item to optimise and test the performance of the hybrid template-machine-learning \photoz code \delight on a narrowband survey; 
    \item to develop an optimal method to calibrate the fluxes between the COSMOS broadbands and the PAUS narrow bands;
    \item to provide an independent \photoz solution for PAUS, enabling the study of photometric and spectroscopic redshift outliers.
\end{enumerate}

This paper is structured as follows. In Section~\ref{sec:data} we first introduce PAUS and the sources of photometry and spectroscopic redshifts used in this work. Section~\ref{sec:algor} describes the algorithms (\delight, \annztwo and \bpz) used in this work, together with their optimisation settings and SED templates used. Section~\ref{sec:calibration} describes the full details of how the photometry and spectroscopy from PAUS, COSMOS and zCOSMOS are cross-matched, how the galaxy fluxes are selected, the three methods to calibrate the broadband and narrowband fluxes, and the performance metrics used in this work to compare the results between runs and codes. Section~\ref{sec:res_delight} shows the \photoz results obtained by \delight, and a thorough analysis is conducted to compare its performance with \annztwo, \bpz and \bcnz. Finally, in Section~\ref{sec:app} we study the \photoz outliers of \delight and \bcnz, and derive new metrics with improved \photoz outlier identifications. Our work is concluded in Section~\ref{sec:conc}.

\section{Photometry and Spectroscopy}\label{sec:data}
In this work, photometric data were obtained from PAUS (Section~\ref{sec:pau}) and \mbox{COSMOS} (Section~\ref{sec:cosmos}), while spectroscopic redshifts were obtained from \mbox{zCOSMOS} (Section~\ref{sec:zcosmos}). In this section, these surveys will be introduced, together with the selection cuts used to obtain our training and testing sets. 

\subsection{PAUS}\label{sec:pau}
PAUS is a narrowband photometric galaxy survey aimed at mapping the large-scale structure of the Universe up to $i\sim23.0$. Using $40$ narrow bands spaced by $100$~\r{A} in the range between $4500$ to $8500$~\r{A} \citep[filter responses visualised in][and Fig.~\ref{fig:overlapbands}]{eriksen_pau_2019}, PAUS aims to achieve redshifts with a precision of $\sigma_\rf{RMS}<0.0035(1+z)$ for galaxies with $i_\rf{auto}<22.5$. PAUS uses the PAUCam instrument \citep{padilla_physics_2019} on the $4$ m William Herschel Telescope (WHT) at Observatorio del Roque de los Muchachos (ORM) in La Palma. It has observed more than $50$ deg$^2$ of sky since the beginning of $2016$, and observations to full depth in all narrow bands for $100$ deg$^2$ are planned.

The PAUS forced-aperture coadded photometry has its aperture defined by using the $50$ per cent light radius ($r_{50}$), the point spread function (PSF), ellipticity and S\'ersic index of COSMOS morphology, such that the fluxes measure a fixed fraction of light. The reader is referred to \citet{eriksen_pau_2019} for detailed information on how the PAUS fluxes are measured. In this work we used the early data release from PAUS (objects are observed at least five times, using an elliptical aperture with $62.5$ per cent light radius), and select objects with $i_\rf{auto}\le 22.5$, entries with no missing measurement, and the COSMOS flag \texttt{TYPE=0} (extended objects).

\subsection{COSMOS}\label{sec:cosmos}
The Cosmic Evolution Survey \citep[COSMOS,][]{scoville_cosmic_2007} covers a sky area of $2$ deg$^2$ ($149.47^{\circ}\le\alpha\le150.7^{\circ}$, $1.62^{\circ}\le\delta\le2.83^{\circ}$) and is known for its high sensitivity, depth and an exceptionally low and uniform Galactic extinction ($E_{B\text{-}V}\sim 0.02$).

In this work we used photometry from the COSMOS$2015$ Catalogue \citep{laigle_cosmos2015_2016}; it is a highly complete mass-selected sample to very high redshifts, highly optimised for the study of galaxy evolution and environments in the early Universe. The COSMOS$2015$ Catalogue provides $30$ band photometry ranging from near UV to near infrared wavelengths, all these have been observed through multiple facilities, two of which are the Canada-Hawaii-France Telescope (CFHT) and Subaru Telescope \citep{miyazaki_subaru_2002}. From this catalogue we only use the CFHT $u^*$-band \citep{boulade_megacam:_2003} and Subaru $B$, $V$, $r$, $i^+$ and $z^{++}$ bands \citep{miyazaki_subaru_2002}, in conjunction with the narrowband photometry of PAUS. For simplicity, these bands will be referred to collectively as the $uBVriz$ bands;  the superscripts are dropped for easier reading.

\subsection{zCOSMOS}\label{sec:zcosmos}
The zCOSMOS Survey \citep{lilly_zcosmos:_2007} targets galaxies in the COSMOS field using the Visible Multi-Object Spectrograph \citep[VIMOS,][]{le_fevre_commissioning_2003}. zCOSMOS-Bright observed $20\,689$ galaxies in a sky area of $1.7$~deg$^2$, these galaxies have magnitudes $15<i_\rf{auto}<22.5$ and redshifts in the range of $0.1<z<1.2$, its spectral range is in the red (rest-frame wavelength $5550$~\r{A} to $9650$~\r{A}) to follow strong spectral features around the $4000$~\r{A} break to as high redshifts as possible.

In this work we use data from zCOSMOS-Bright DR$3$\footnote{\url{http://www.eso.org/qi/catalog/show/65}}. Galaxies with redshift confidence class $3$ and $4$ (spectroscopic verification rate of $99$ and $99.8$ per cent, respectively) are selected and cross-matched with PAUS objects.

\subsection{Our dataset}\label{sec:dataused}
Using the aforementioned selection cuts, we cross-matched within $1''$ the $40$-narrowband photometry from PAUS, six broadband photometry ($uBVriz$) from COSMOS, and highly reliable redshifts from zCOSMOS to obtain a data sample of $8406$ galaxies, which is divided randomly into half for training and testing respectively. This sample uses a total of $46$ bands, and flux calibration between the broad and narrow bands is required as they are obtained from different surveys with different flux measurements. The calibration between these fluxes will be discussed in Section~\ref{sec:calibration}. 

\begin{figure} 
\centering
\includegraphics[width=\linewidth]{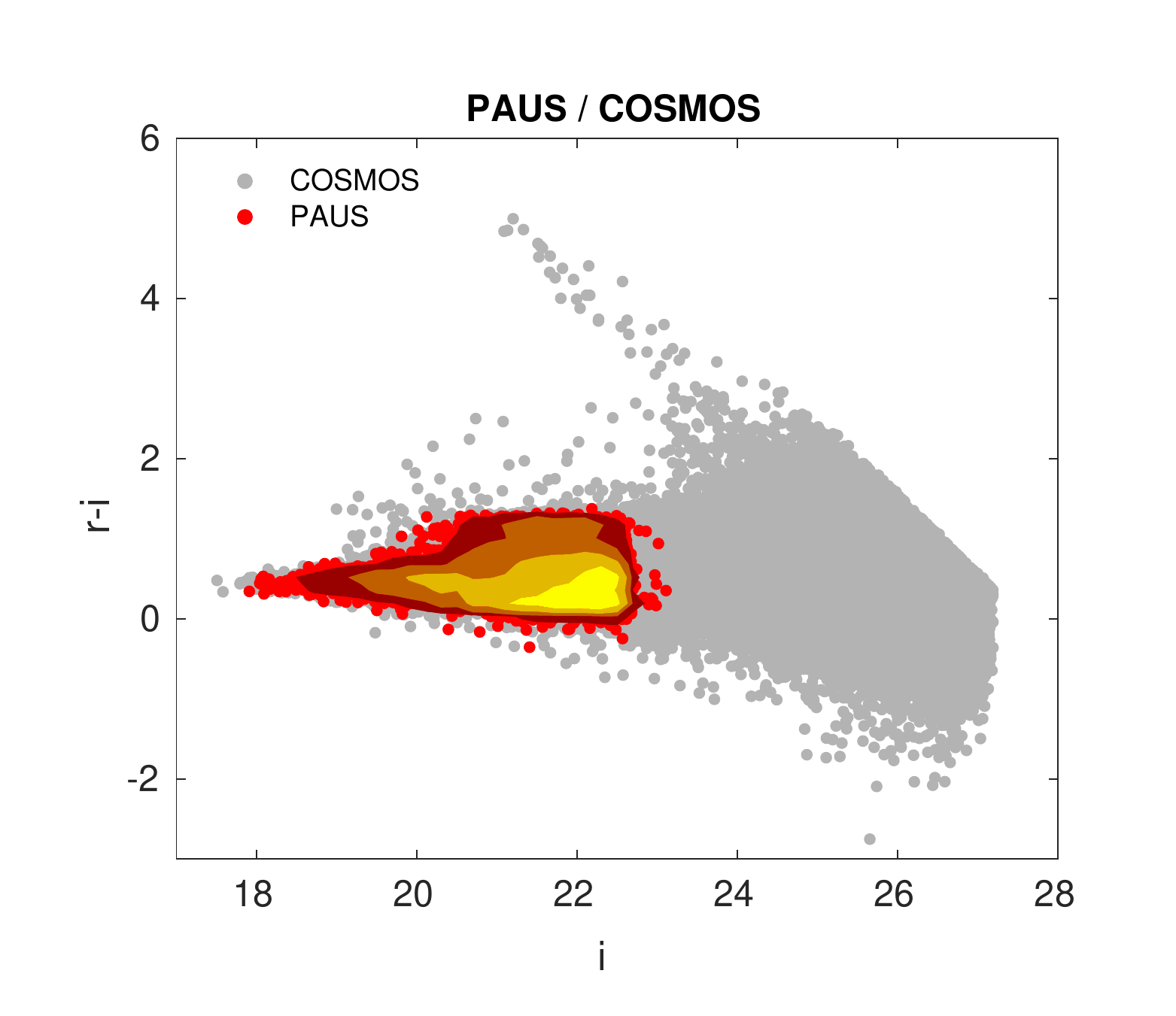}
\caption{Colour-magnitude diagram for the PAUS data (red) used in this work in comparison with the COSMOS2015 sample (all objects with \texttt{TYPE=0} and detected in $r$ and $i$). The contours represent the density of objects.} \label{fig:data}
\end{figure}

The colour-magnitude diagram of this sample is shown in Fig.~\ref{fig:data}, in comparison with the COSMOS$2015$ sample (all objects with \texttt{TYPE=0} and detected in $r$ and $i$). The slight incompleteness in $i$ magnitude is due to the selection effects in brightness of the spectroscopic redshifts available.

The sample size may seem small, but is sufficient for the GP to work, since the GP essentially creates $4000+$ flux-redshift `templates' to produce \photoz's for objects in the testing set. However, we note that such a small training size has a major effect on the results of \annztwo as this training size is close to the lower limit threshold suggested by \citet{bonfield_photometric_2010}. We also note that the sample we have chosen is very similar to that of \citet{eriksen_pau_2019}, the only difference being that they have a more relaxed cut in the number of bands (\tf{N\_BANDS}), being \tf{35<N\_BANDS<40} (workable for a template code like \bcnz), while we used \tf{N\_BANDS=40}\footnote{The relaxed cut resulted in \citet{eriksen_pau_2019} having a larger sample size of $10\,801$ objects.}. When comparing results between \delight and \bcnz, we will only compare \photoz's of the exact same objects. Note that we have used the same broad bands as used by \citet{eriksen_pau_2019}.

\section{Algorithms and Templates}\label{sec:algor}

\subsection{Delight and Gaussian processes}\label{sec:delight}
\delight\footnote{\url{https://github.com/ixkael/Delight}} \citep{leistedt_data-driven_2017} is a hybrid template-based and machine learning \photoz algorithm, which was constructed to combine the advantages, and minimise the disadvantages, of both types of algorithms. \delight constructs a large collection of latent SED templates (or physical flux-redshift models) from training data, with a template SED library as a guide to the learning of the model. This conceptually novel approach uses Gaussian processes (GPs) operating in flux-redshift space. \delight was featured in the results of the LSST \Photoz Data Challenge $1$ \citep{schmidt_evaluation_2020}, where it was found to have a low \photoz bias but slightly broader PDFs.

\begin{figure} 
\centering
\includegraphics[width=1.0\linewidth]{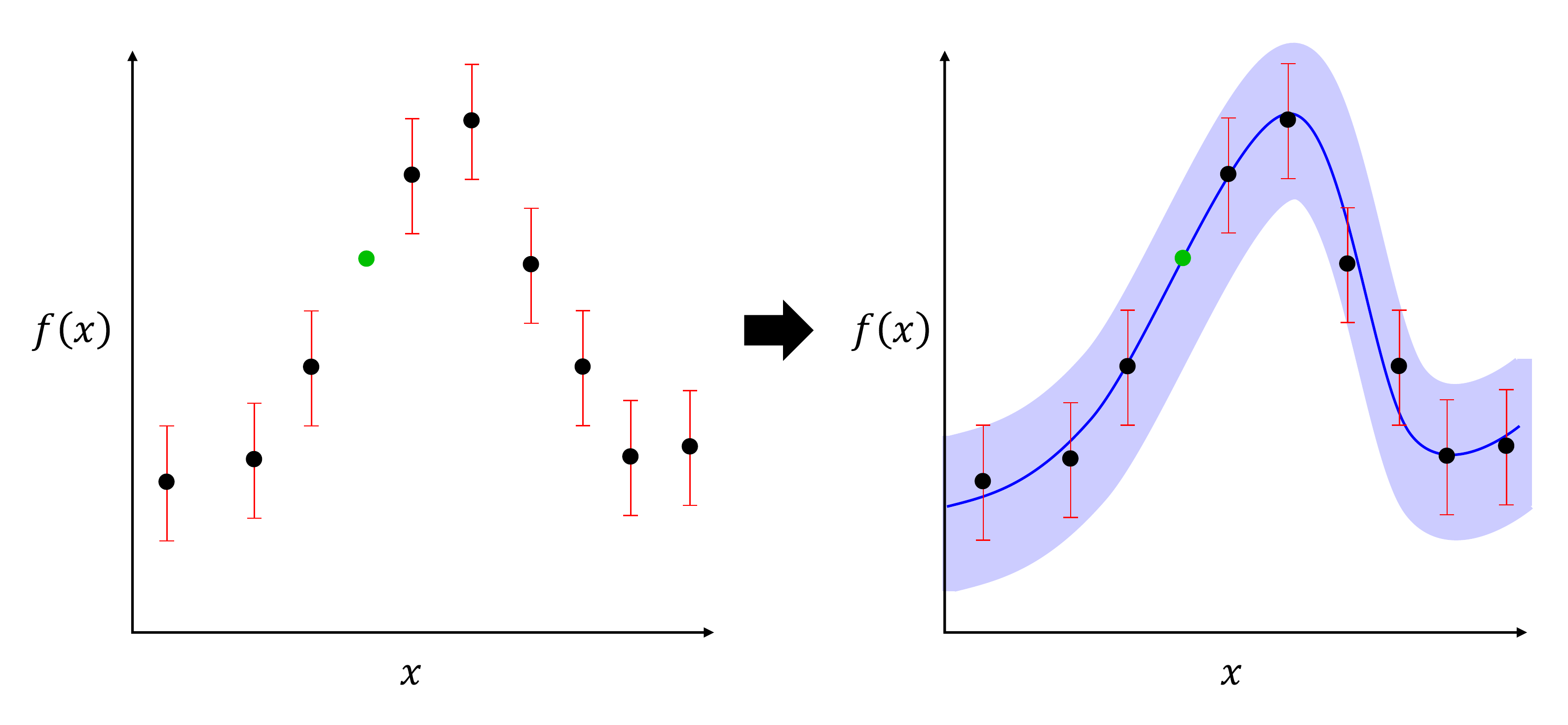}
\caption{Illustration of a Gaussian process (GP). The left panel shows data points (black dots), with a single datum to be predicted (green dot). The GP trains on the given data points to provide a best fit function (blue line) as shown on the right. It also provides a Gaussian confidence interval (blue shaded area) for the prediction.} \label{fig:gp}
\end{figure}

A GP is a supervised learning method, which finds a distribution over the possible functions $f(x)$ that are consistent with the observed data $x$. Consider Fig.~\ref{fig:gp}: suppose we have a set of observed variables $y=f(x)$, we can fit it using a GP, denoted as $f\sim \mathcal{GP}\left(\mu,k\right)$ , which assumes that the probability of all $f(x)$ is jointly Gaussian and representable by a mean function $\mu(x)$ and a covariance matrix $\Sigma(x)=k(x_i,x_j)$. $k(x_i,x_j)$ is the kernel function, which relates one variable $x_i$ to another $x_j$. An example case would be $\mu \equiv 0$ and a kernel function that takes the form of a squared exponential,
\begin{equation}
k(x_i,x_j)=\sigma^2_f \exp\left[ \frac{-(x_i-x_j)^2}{2l^2}\right],
\end{equation}
where $\sigma_f^2$ is the maximum allowable covariance between data (set by the errors on the observation), and $l$ is the tunable correlation length that determines the smoothness of the GP. In this simplistic case, the GP will try to find a marginalisation of all possible functions, but $\mu$ and $k$ can be modified if an underlying model of the data we want to fit is known. The covariance function is defined such that a smooth function is to be predicted. 

Assuming that we have a set of training data $\{x_i,f(x_i)\}$ and would like to find the prediction $\{x_*,f_*(x_*)\}$, the GP models $f$ and $f_*$ as jointly Gaussian, $\mathcal{N}(\mu,\Sigma)$, and therefore
\begin{equation}
\left( \begin{array}{cc} f(x) \\ f_{*}(x) \end{array} \right) \sim \mathcal{N}\left( \left( \begin{array}{cc} \mu \\ \mu_{*} \end{array} \right), \left( \begin{array}{cc}
\Sigma & \Sigma_{*} \\ \Sigma_{*}^T & \Sigma_{**} \end{array} \right) \right),
\end{equation}
where $\Sigma=k(x_i,x_j)$ is the covariance between the training data, $\Sigma_*=k(x_*,x_i)$ the covariance between training and the predicted data (superscript $T$ denotes the transpose of the matrix), while $\Sigma_{**}=k(x_*,x_*)$ is the variance of the predicted data. 

It follows from the above that the posterior $p(f_*|x_*,x_i,f_i)$ is also Gaussian, therefore a predicted point $f_*(x_*)$ is plotted (green dot in Fig.~\ref{fig:gp}) is modelled by a Gaussian function (smooth blue line) which runs across all points, with its $95\%$ confidence interval ($\pm1.96\sigma_{f_*}$) represented by the navy shaded area.

In the context of \delight, GPs are used to calculate the predicted fluxes $\hat{F}$ at a certain redshift $z$ for a training object $i$ with fluxes $F_i$ and redshift $z_i$. This could be better understood by first defining the posterior \photoz distribution $p(z|\hat{F})$ of an object in the testing set. For machine learning methods, it has the form
\begin{equation} \label{eq:algor:pzf1} 
p(z|\hat{F}) \approx \sum_i p(\hat{F}|z,z_i,F_i)\, p(z|z_i,F_i)p(z_i,F_i),
\end{equation}
where $p(\hat{F}|z,z_i,F_i)$ is the prediction for fluxes of the training galaxy at a different redshift $z$, while $p(z|z_i,F_i)$ and $p(z_i,F_i)$ are the priors that provide the redshift distributions and abundances, generated from the training data, which are multiplied to give the combined probability $p(z,z_i,F_i)$ for a given redshift $z$ and training object with redshift $z_i$ and fluxes $F_i$. This is analogous to the one derived from template-based methods,
\begin{equation} \label{eq:algor:pzf2} 
p(z|\hat{F}) \approx \sum_i p(\hat{F}|z,t_i)\, p(z|t_i)p(t_i),
\end{equation}
where $t_i$ is the template, $p(z|t_i)p(t_i)=p(z,t_i)$ is the prior and $p(\hat{F}|z,t_i)$ is the probability of the predicted flux $\hat{F}$ at redshift $z$ and for template $t_i$. Both equations are easily differentiated by the fact that for template-based methods, $p(z|\hat{F})$ is derived using a list of templates $t_i$, while for machine learning methods it is derived using the individual training set objects with fluxes $F_i$ and spectroscopic redshift $z_i$. 

\delight differs a little from the usual machine learning method in the sense that instead of finding a direct empirical relationship between the fluxes and redshifts of the training objects, it uses a GP to model the predicted fluxes of a training galaxy at different redshifts with the help of SED templates. This creates a latent flux-redshift template for each training object, where for a given set of fluxes in the testing set, it could be compared to several training templates to find the best predicted redshift. 

The algorithm first fits a best-fit SED template to a particular training object $i$ with redshift $z_i$ and fluxes $F_i$ (multiple bands); the best-fit SED template is then used to formulate the mean function and kernel of a GP to build a flux-redshift template which could predict the expected fluxes of certain band filters when this object is redshifted to a different $z$. With each training object now becoming a flux-redshift template, the final \photoz posterior distribution of a testing set object is determined by making a pairwise comparison of every training-testing pair, and a weighted solution is obtained based on the best fits of each pair. 

In other words, we are computing the probability that the target galaxy has the same SED as the training galaxy but at a different redshift. \delight is thus a hybrid template-machine-learning \photoz algorithm in the sense that SED templates are used to `guide' the creation of flux-redshift templates based on the training objects, or, if seen from another perspective, the GP `corrects' the SED templates by using training data. We refer the reader to \citet{leistedt_data-driven_2017} for more on Gaussian processes, and also for the full expressions of the $\mu$ and $k$ in relation to the filter responses, flux normalisations, linear mixtures of physical SED templates, and the manually configurable SED residual function of emission lines.

\delight is advantageous over many other \photoz algorithms as its output is less dependent on representative training data, and it does not strictly require the training set to use the same photometric bands. However, it still requires accurate spectroscopic redshifts, high quality training fluxes and representative templates to produce high quality \photoz probability density functions (PDFs), or $p(z)$. As such, given a few photometric bands, \delight is able to predict missing bands or fluxes in an entirely different set of photometric bands, and this function is utilised in Section~\ref{sec:calibpergal} to predict and calibrate the flux values between two surveys.

\subsection{Delight optimisation} \label{algor:delight:optim}
The optimisation settings of \delight used in this work are as follows. For the GP setup, the number of Gaussians to fit the filter curves (\tf{numGpCoeff}) was set to $7$ instead of the default $20$, appropriately selected to accommodate the smaller full width half maximum (FWHM) of the narrowband filters. Other than that, we have mainly used the default hyperparameter settings for \delight with the exception of the widths of the luminosity and redshift priors $\sigma_\ell$ and $\sigma_z$ \citep[\tf{ellPriorSigma} and \tf{zPriorSigma}, see][]{leistedt_data-driven_2017}, which have been lowered to $0.2$ and $0.1$ respectively as they produced better results. 

As mentioned earlier, the mean function and the kernel of the GP are modelled after the choice of emission lines and SED template sets. We replaced the $3$ default emission lines in \delight with the list provided by \citet{eriksen_pau_2019}, although we note that the change in result for this is insignificant. As for the templates, we used the \citet{brown_atlas_2014} high-quality templates, which consist of $129$ SEDs derived from real nearby galaxies. These templates have wavelengths covering the ultraviolet to mid-infrared, and encompass a broad range of galaxy types including ellipticals, spirals, merging galaxies, blue compact dwarfs and luminous infrared galaxies. In this work we have also tested the performance of various other template sets \citep{coleman_colors_1980,kinney_template_1996,bruzual_stellar_2003,ilbert_accurate_2006,polletta_spectral_2007}; however they do not perform as well as those of \citet{brown_atlas_2014}: the root-mean-square \photoz errors could range between $21$ to $112$ per cent higher when these templates are used. Therefore, the results from these tests are not shown in this work.

We note that \delight requires all magnitudes $m_i$ and magnitude errors to be converted into fluxes $F_i$ and flux variances, with a zero-point adjustment of $26.4$ in magnitude (i.e. $F_i=10^{-0.4(m_i-26.4)}$). We have also added a $3$ and $6$ per cent flux error in quadrature to the flux variances for the narrow and broad bands, respectively, to account for other flux errors from both the data and the model (values estimated via trial and error). It is also worth mentioning that while \delight is capable of processing negative fluxes (non-detections), the reference band (\tf{referenceBand}) used for flux normalisation  only handles fluxes with positive values. In this work we have selected the narrow band $nb625$ as the reference band, or the COSMOS $r$-band in cases where narrow bands were not used. 

Throughout this work, we use $z_\rf{map}$ (the \textit{maximum a posteriori} of the PDF) to represent the best point estimate \photoz produced by \delight. The output \photoz PDF bins were set to be linear instead of logarithmic, with a stepsize of $0.001$, and a range of $0.02<z<1.65$, keeping close to the limits of the spectroscopic redshifts.

\subsection{Other algorithms}\label{sec:annz2bpz}
We are also interested in how \delight compares to other common template-based or machine-learning-based methods besides \bcnz and \deepz. Therefore two other \photoz algorithms, \annztwo and \bpz are also used in this work, using the same training and template sets, to be compared with the performance of \delight. In the following paragraphs we briefly introduce the two algorithms and their optimisation settings.

\annztwo \footnote{\url{https://github.com/IftachSadeh/ANNZ}} \citep{sadeh_annz2:_2016} is a machine-learning-based \photoz algorithm which has been widely used in recent works \citep{bonnett_redshift_2016,jouvel_photometric_2017,bilicki_photometric_2018,soo_morpho-z:_2018,schmidt_evaluation_2020} due to its high customisability and its ability to produce PDFs. It uses the Toolkit for Multivariate Data Analysis \citep[TMVA,][]{hoecker_tmva_2007} with \textsc{root} \citep{brun_root_1997}, which allows it to run multiple different machine learning algorithms for training, and outputs \photoz's based on a weighted average of their performance. In this work we ran \annztwo with a mixture of $3$ machine learning methods, namely artificial neural networks (ANNs), boosted decision trees (BDTs) and $k$-nearest neighbours (KNNs), see \citet{hoecker_tmva_2007} for detailed descriptions of these machine learning algorithms. An architecture of $N$:$\frac{2N+1}{3}$:$\frac{N+2}{3}$:$1$ was used for the ANN; the bagging method was used to boost the decision trees; a polynomial kernel was used for the KNN; while the other hyperparameters for each method were individually optimised for best performance. \annztwo version $2.3.1$ was used in this work, and the mean value of the PDF, $z_\rf{pdf}$ was chosen to represent the \photoz point estimate.

\bpz\footnote{\url{http://www.stsci.edu/~dcoe/BPZ/}} \citep{benitez_bayesian_2000}, on the other hand, is one of the longest-standing template-based \photoz algorithms, and still widely used today \citep{marti_precise_2014,bundy_stripe_2015,cavuoti_cooperative_2017,tanaka_photometric_2018,joudaki_kidsviking-450_2020,raihan_testing_2020}. Other than sharing the usual attributes of a template-based code, \bpz uses Bayesian inference, prior information of redshift distributions and template interpolation to improve \photoz results. \bpz version $1.99.3$ was used in this work, and similar to \delight the Brown templates were used, with the interpolation parameter set to $2$. We assumed the same functional form for the Bayesian priors as those used by COSMOS \citep{laigle_cosmos2015_2016}. The peak of the PDF, $z_\rf{b}$, was used as the best \photoz point estimate. 

Other than \annztwo and \bpz, the results of \delight are also compared to the results of \bcnz, which was developed specifically for the PAUS data \citep{eriksen_pau_2019}. \bcnz is able to compute a linear combination of SED templates and  is designed to deal with emission lines, extinction, and adjust zero-points between narrow and broad bands, all of which are crucial in the context of PAUS. The introduction of the code \bcnz and its early demonstration of PAUS \photoz can be found in \citet{eriksen_pau_2019}.

\section{Flux Calibration}\label{sec:calibration}
This work utilises fluxes obtained from two different surveys: the PAUS narrowband fluxes are measured using an aperture which covers $62.5$ per cent of light from the galaxy, while COSMOS broadband fluxes are measured using a fixed $3''$ aperture. Therefore, calibration is required to ensure that the flux values are consistent with one another. We only calibrate the broadband fluxes, leaving the narrowband fluxes untouched following \cite{eriksen_pau_2019}. The calibration process is done in two steps: first we derive empirical corrections to account for differences in the aperture photometry (calibration for each galaxy), then placing all bands at the same flux zero point (calibration for each band). For the correction for differences in flux aperture, we note that ideally this could have been easily done if \specz's are available; however since the evaluation set would not have \specz's available, we present $3$ alternatives in the following sections to calibrate the fluxes photometrically.

\subsection{Correction for differences in flux aperture}\label{sec:calibpergal}
In the first step we define a parameter $R_\rf{g}$, a correction factor estimated for each galaxy to be multiplied with all of its six $uBVriz$ broadband fluxes. Ideally, this factor is estimated by first finding the best-fit Brown template for each galaxy using only $40$ narrowband fluxes from PAUS and its true redshift. The best-fit template is then used to generate the predicted $uBVriz$ fluxes, and a weighted mean of the ratios between the predicted flux and the original COSMOS flux $R_{\rf{g},b}$ is calculated for each band $b$, given by
\begin{equation}
R_\rf{g} = \frac{ \sum_b R_{\rf{g},b}/\sigma_{R_\rf{g},b}^2 }{ \sum_b 1/\sigma_{R_\rf{g},b}^2 },
\end{equation}
where the sum is over the six COSMOS broad bands, and $\sigma_{R_\rf{g},b}^2$ is the variance of $R_{\rf{g},b}$. Here we have assumed that the Brown templates are sufficiently representative, and therefore the predicted flux derived from it is the true flux of the broad bands. We have also assumed that $R_{\rf{g},b}$ should be almost the same across each band for each galaxy. This calibration is motivated by the fact that each galaxy requires a calibration between fixed-size and adaptive aperture photometry dependent on its apparent size. 

We now explore three different methods to determine $R_\rf{g}$ from the photometric data only.

\subsubsection{The photo-z calibration method}\label{sec:photozmethod}
The first method, which we call the \textit{\photoz calibration method}, is very similar to the method above except that we replace the spectroscopic redshifts used to determine the predicted $uBVriz$ flux for the testing set with photometric redshifts. We first use \delight and only the $40$ narrow bands to produce \photoz's for each object, and then we use these \photoz's to estimate the predicted fluxes, and then later $R_\rf{g}$ for each galaxy. This implies that the better the quality of the \photoz's produced by only the $40$ narrow bands, the better the calibrated broadband fluxes will be.

\subsubsection{The size calibration method} \label{sec:sizemethod}
The second method, hereafter the \textit{size calibration method}, does not require the production of predicted fluxes for the testing set. Instead, this method uses the correlation between the sizes of galaxies with their values of $R_\rf{g}$ in the training set, to predict the values of $R_\rf{g}$ for objects in the testing set. With the predicted fluxes of the training set known, we plot $R_\rf{g}$ against the $50$ per cent light radius $r_{50}$ (measured in pixels) for each object, and obtain a best-fit linear-least-squares regression line in the process,
\begin{equation}
R_\rf{g} = m\cdot r_{50} + c,    
\end{equation}
where the slope and $y$-intercept are found to be $m=0.0101$ and $c=0.4504$ respectively, with a correlation coefficient of $r=0.8349$, implying a strong positive correlation between $R_\rf{g}$  and $r_{50}$.

\begin{figure} 
\centering
\includegraphics[width=0.83\linewidth]{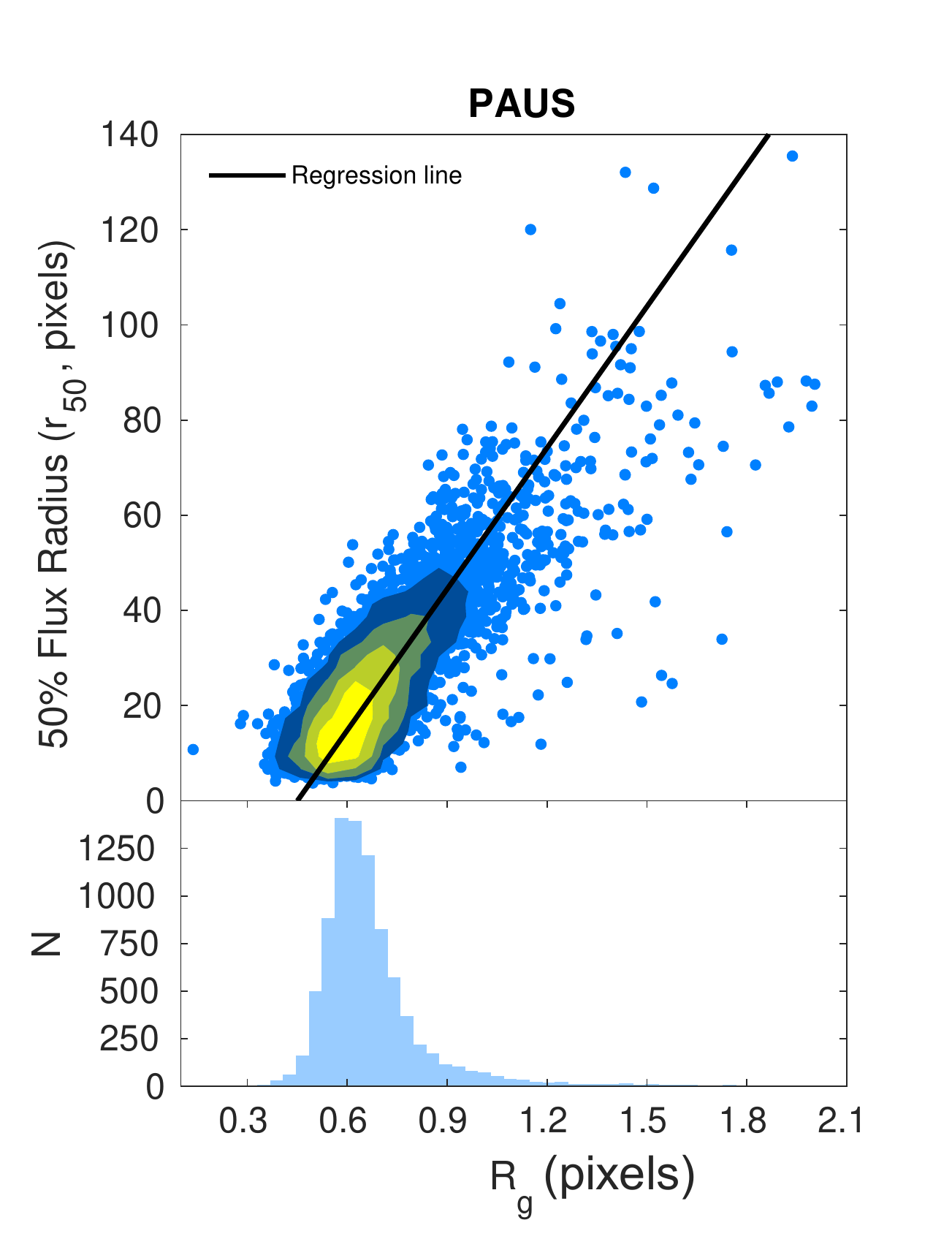}
\caption{\textit{Top}: Correlation between $r_{50}$ and $R_\rf{g}$ for the training set, where $R_\rf{g}$ is a calibration correction factor estimated for each galaxy to be multiplied with all of its six $uBVriz$ broadband fluxes. \textit{Bottom}: The distribution of $R_\rf{g}$ of the training set, estimated using the size calibration method. $N$ is the number of galaxies.} \label{fig:rgr50}
\end{figure}

With this relationship derived, the values of $R_\rf{g}$ for each object in the testing set can be estimated. This method is motivated by the fact that the size of galaxies is a defining factor for the difference in their flux values when measured using a fixed aperture or when measured using a fixed light radius. Fig.~\ref{fig:rgr50} shows a scatter plot of $r_{50}$ v.s. $R_\rf{g}$ for the training set, where the correlation equation is determined. The distribution of $R_\rf{g}$ is also tabulated in the figure, it is shown to have a median value of $0.6349$, implying that on average COSMOS measures more flux for each galaxy than PAUS. We note that in the case when galaxies have undefined values of $r_{50}$, we substitute them with the mean value of $r_{50}=22.4934$~pixels.

\subsubsection{The flux calibration method} \label{sec:fluxmethod}
\begin{figure*} 
\centering
\includegraphics[width=\linewidth]{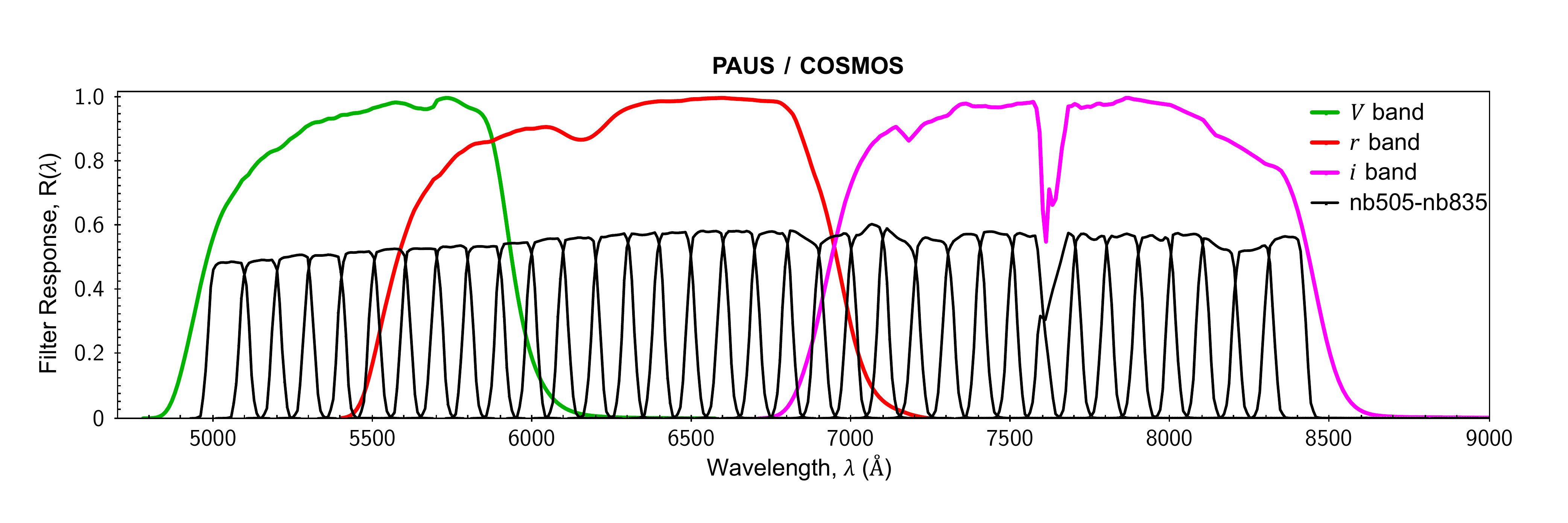}
\caption{The overlapping wavelengths between $34$ PAUS narrowband filters and $3$ COSMOS broadband filters: $V$ overlaps with $nb505$-$nb585$ ($9$ bands); $r$ overlaps with $nb565$-$nb685$ ($13$ bands); and $i$ overlaps with $nb705$-$nb835$ ($13$ bands). Note that the filter responses from PAUS and COSMOS are normalised at different values, respectively.} \label{fig:overlapbands}
\end{figure*}

The third and final method is the \textit{flux calibration method}, which is similar to the method used by \citet{eriksen_pau_2019}, but simpler in that the Gaussian Process has a larger capacity to accommodate uncertainties. This method makes use of the fact that there are overlaps in wavelength between the COSMOS broad bands and PAUS narrow bands: the $V$-band overlaps with the narrow bands $nb505$ to $nb585$ ($9$ bands); the $r$-band overlaps with $nb565$ to $nb685$ ($13$ bands); and the $i$-band overlaps with $nb705$ to $nb835$ ($14$ bands). This overlap is illustrated in Fig.~\ref{fig:overlapbands}.

Similar to the previous method, no redshift information is required for flux prediction, the $R_\rf{g}$ in this case is estimated by first averaging the narrowband fluxes within the range of the broad band of interest ($V$, $r$ or $i$), and then taking the ratio between the broadband flux and the averaged narrowband fluxes. This will give us $3$ values of $R_{\rf{g},b}$ for the $3$ $Vri$ bands, and finally $R_\rf{g}$ for each galaxy is taken as the weighted average of the $3$ values. 

This method is simple yet effective: it does not involve the spectroscopic redshift, the \photoz derived by $40$ narrow bands, or even the size of the galaxy. Here we assume that the $R_\rf{g}$ estimated using $Vri$ is applicable for the $uBz$ bands as well. We will compare the overall \photoz quality produced by the three methods above in Section~\ref{sec:quality}.

\subsection{Correction to flux zero points}\label{sec:calibperband}
After calibrating the COSMOS broadband fluxes for each galaxy, we proceed to calibrate the broadband magnitude offsets within each band. We perform a weighted least-squares fit between the predicted broadband fluxes (produced by \delight using $40$ PAUS narrowband fluxes, the respective best fit Brown templates and zCOSMOS \specz's) and the original COSMOS $uBVriz$ fluxes in the training set, by using a simple linear equation,
\begin{equation}\label{eq:lnF}
\ln(F_{\rf{p},b}) = a_b \cdot \ln(F_{\rf{g},b})+c_b
\end{equation}
where $F_{\rf{p},b}$ is the predicted flux for band $b$, $F_{\rf{g},b}$ the COSMOS broadband flux after undergoing the per-galaxy calibration, and $a_b$ and $c_b$ are constants to be optimised. The values of $a_b$ and $c_b$ estimated for each band using the training set are now used to calibrate the fluxes in the testing set, and these values are tabulated in Table~\ref{tab:calibband}. A weighted fit was implemented, with the inverse variances of the fluxes used as the weights, since we expect that objects which are brighter to have relatively lower variances, and by accounting for the variances of objects the fainter objects would be upweighted. 

\begin{table} 
\centering
\small
\caption{List of the best fit parameters $a_b$ and $c_b$ for each band $b$ when the predicted and original COSMOS fluxes from the training set were fitted with a weighted least-squares fit, using Equation~\ref{eq:lnF}.} \label{tab:calibband}
\begin{tabular}{lrr}
\hline 
\textbf{Bands}  & \textbf{$a_b$}    & \textbf{$c_b$} \\
\hline 
$u$           & $1.0007\pm0.0001$ &  $0.0354\pm0.0008$  \\
$B$             & $0.9906\pm0.0002$ &  $0.2163\pm0.0009$    \\
$V$             & $0.9988\pm0.0002$ & $-0.0830\pm0.0009$   \\
$r$             & $1.0006\pm0.0002$ &  $0.0015\pm0.0009$ \\
$i$           & $1.0202\pm0.0001$ & $-0.0875\pm0.0008$  \\
$z$        & $0.9791\pm0.0001$ &  $0.0424\pm0.0007$ \\
\hline 
\end{tabular}
\end{table}

As expected from the table, the values of $a_b$ and $c_b$ are very close to $1$ and $0$ respectively, since the calibrated flux for aperture correction $F_{\rf{g},b}$ is already very close to the predicted flux $F_{\rf{p},b}$. Essentially, this process `straightens' the correlation line, providing minor yet essential improvements to the overall calibration.

\subsection{Overall calibration performance} \label{sec:caliboverall}

\begin{figure*} 
\centering
\includegraphics[width=\linewidth]{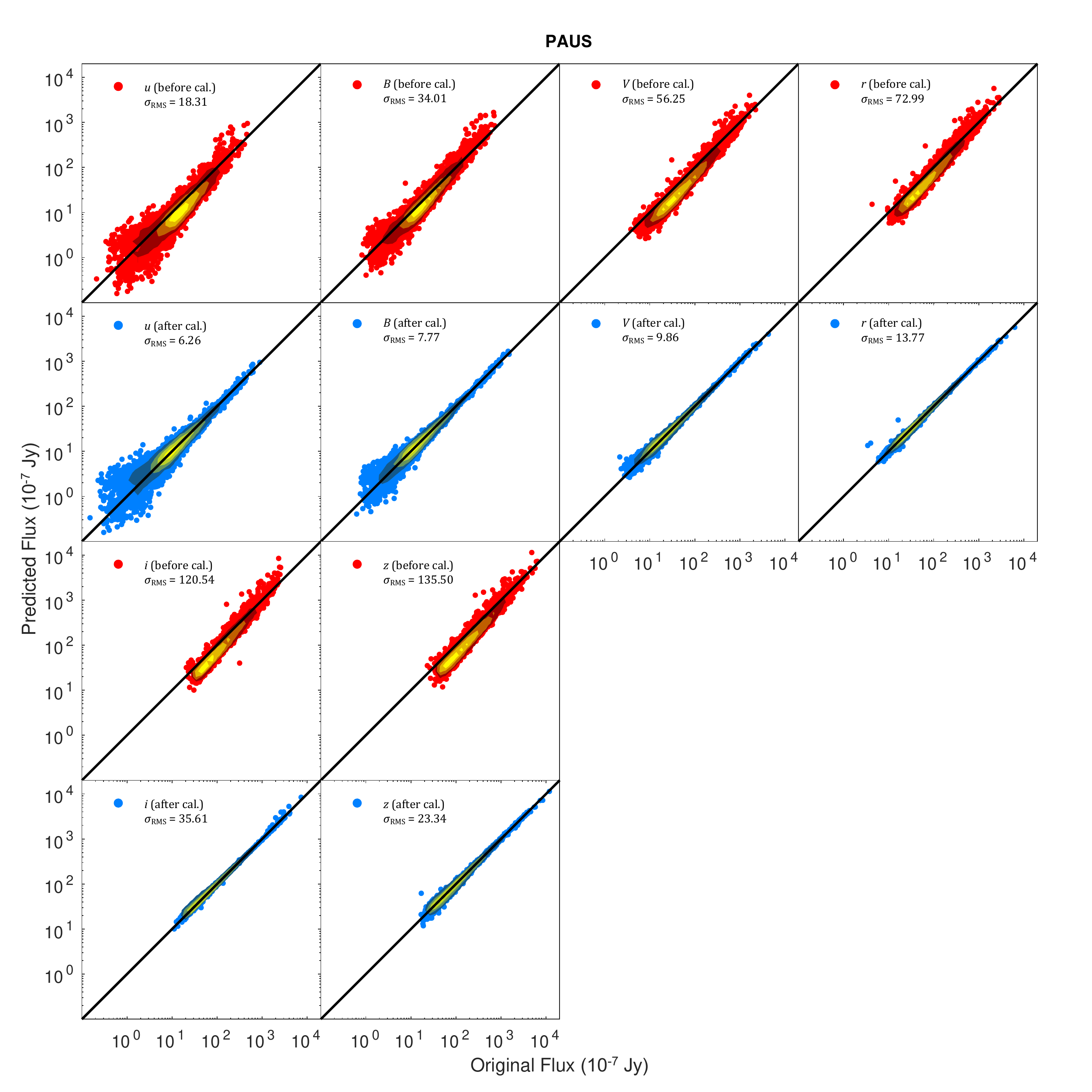}
\caption{The $uBVriz$ broadband fluxes predicted by \delight plotted against their original COSMOS fluxes, both before and after the two-step calibration process (red and blue respectively) for our training set, using the flux calibration method as an example. Based on the root-mean-square errors ($\sigma_\rf{RMS}$) shown in each panel, the broadband fluxes match their prediction much better after calibration.} \label{fig:calibresults}
\end{figure*}

Figure~\ref{fig:calibresults} shows the correlation between the broadband fluxes predicted by \delight (using spectroscopic redshifts, PAUS $40$ narrow bands and Brown templates) and the COSMOS broadband fluxes for our training set, both before and after calibration (red and blue, respectively). The figure only shows the result of the flux calibration method, as the other two methods look very similar graphically (which translates to a small difference in \photoz results shown later in Section~\ref{sec:res_delight}). 

The RMS values displayed in Figure~\ref{fig:calibresults} show that for all bands, the scatter between the original fluxes with respect to the predicted fluxes has reduced by $63$ to $88$ per cent after the two-step calibration was done. The scatter at low fluxes for the $u$ and $B$-bands remains evident, which originated from the high uncertainty in flux measurements. Despite the large decrease in scatter, we note that the RMS value here is not a metric of improvement for calibration as we do not have the true values of the broadband fluxes in the matched apertures. However, the calibration of the broadband fluxes did translate into an improvement in \photoz scatter and $68$th percentile error by about $70$ to $80$ per cent, as shown in Section~\ref{sec:res_delight}.

\section{Results and Discussion}\label{sec:res_delight}

\begin{table*}  
\caption{The root-mean-square error ($\sigma_\rf{RMS}$), $68$th percentile error ($\sigma_{68}$), outlier fraction ($\eta_\rf{out}$), mean continuous ranked probability score ($\rho_\rf{CRPS}$) and the root-mean-square error in redshift distribution ($n_\rf{RMS}$) for the \photoz's produced in this work, using different algorithms, methods and number of bands. All results are produced using $6$ broad bands (BB) and $40$ narrow bands (NB) unless stated otherwise.} \label{tab:results}
\begin{tabular}{lrrrrr}
\hline
\textbf{\Photoz methods} & $\sigma_\rf{RMS}$ & $\sigma_{68}$ & $\eta_\rf{out}$ ($\%$) & $\rho_\rf{CRPS}$ & $n_\rf{RMS}$ \\
\hline
\delight ($6$BB only)                       & $0.0514$  & $0.0441$  & $0.93$                 & $0.0388$ & $0.885$   \\
\delight ($40$NB only)                      & $0.0684$  & $0.0119$  & $4.02$                 & $0.0298$ & $0.637$   \\
\hline
\delight (no calibration)                   & $0.1555$  & $0.0566$  & $9.06$                & $0.0887$ & $0.895$    \\
\delight (\photoz calibration method)       & $0.0335$  & $0.0083$  & $0.71$                 & $0.0158$ & $0.634$    \\
\delight (size calibration method)          & $0.0341$  & $0.0095$  & $0.76$                 & $0.0165$ & $0.646$    \\
\delight (flux calibration method)          & $0.0331$  & $0.0081$  & $0.86$                 & $0.0155$ & $0.636$    \\
\delight (flux calibration method, no GP)   & $0.0442$  & $0.0089$  & $0.98$                 & $0.0179$ & $0.639$    \\
\hline
\annztwo                                    & $0.0556$  & $0.0396$  & $2.66$                 & $0.0719$ & $0.465$    \\
\annztwo ($6$BB only)                    & $0.0371$  & $0.0202$  & $1.14$                 & $0.0522$ & $0.432$    \\
\bpz                                        & $0.0368$  & $0.0089$  & $0.86$                 & $0.0184$ & $0.740$    \\
\hline
\bcnz                                       & $0.0403$  & $0.0085$  & $1.14$                 & $-$       & $-$        \\
\hline
\end{tabular}
\end{table*}

Table~\ref{tab:results} summarises the results of this work, it shows all the \photoz metrics we produced, using different algorithms (\delight, \annztwo, \bpz), different calibration methods (\textit{flux}, \textit{\photoz} and \textit{size}), and different number of input fluxes ($6$ broad bands, $40$ narrow bands, or both). We divide the analysis of the results into two sections: Section~\ref{sec:quality} studies the performance between the three calibration methods used in \delight, while Section~\ref{sec:comparison} compares the best performance of \delight with \annztwo, \bpz and \bcnz. In the following section, we briefly introduce the performance metrics we used in this work.

\subsection{Performance metrics}\label{sec:metrics}
In this work we use three metrics to quantify the performance of the \photoz point estimates: the root-mean-square error ($\sigma_\rf{RMS}$), the $68$th percentile error ($\sigma_{68}$) and the outlier fraction rate ($\eta_\rf{out}$). With $\Delta z \equiv \frac{z_\rf{phot}-z_\rf{spec}}{1+z_\rf{spec}}$, the above metrics are defined as follows:
\begin{equation} 
    \sigma_\rf{RMS}\equiv \sqrt{\frac{1}{N} \sum_i^N \left|\Delta z_i\right|^2}\ ,
\end{equation}
\begin{equation}
\sigma_{68} \equiv \frac{Q_{84.1\%}(\Delta z_i)-Q_{15.9\%}(\Delta z_i)}{2}\ ,
\end{equation}
\begin{equation}\label{eq:outlier}
\eta_\rf{out} \equiv \rf{\% objects where} \left| \Delta z_i \right| \geq 0.15\ .
\end{equation}
Here $N$ is the total number of galaxies, while $Q$ is a percentile of the distribution. Since $\sigma_\rf{RMS}$ is calculated without the outliers removed, it measures the overall scatter of the sample, whereas $\sigma_{68}$ measures the scatter with reduced sensitivity to outliers. 

With similar motivations as \citet{marti_precise_2014} and \citet{eriksen_pau_2019}, we hope to achieve an overall \photoz error of $\sigma_{68}\leq 0.0035(1+z_\rf{spec})$ for at least $50$ per cent of the testing sample after applying an appropriately chosen quality cut. We use the Bayesian ODDS ($\Theta$) parameter \citep{benitez_bayesian_2000} in \delight, similar to its implementation in \annztwo by \citet{soo_morpho-z:_2018}. $\Theta$ can be estimated from the \photoz PDF, $p(z)$ using the equation
\begin{equation} 
\Theta=\int^{z_\rf{p}+k(1+z_\rf{p})}_{z_\rf{p}-k(1+z_\rf{p})} p(z)\, dz\ ,
\end{equation}
where $z_\rf{p}$ is the peak of $p(z)$ and $k=0.01$. $\Theta$ ranges between $0$ and $1$, the higher the value the lower the $p(z)$ width, which implies a more precisely predicted \photoz (though not necessarily accurate). The value of $k$ is arbitrary, appropriately selected such that not too many objects end up having $\Theta=1$. Therefore, an $x$ per cent quality cut on the sample keeps the top $x$ percent of objects with the highest values of $\Theta$.

To assess the quality of the $p(z)$, we use probability integral transform (PIT) plots and the continuous ranked probability score (CRPS). The PIT is the cumulative distribution function (CDF) at $z_\rf{spec}$ while asserting the $p(z)$ to have an area of unity. Since the \photoz CDF is $C(z)=\int^z_{0} p(z')\ dz'$, PIT is defined to be
\begin{equation} 
\rf{PIT} = C(z_\rf{spec}) = \int_0^{z_\rf{spec}} p(z) \, dz\ .
\end{equation}
A PIT distribution tells us on average if the $p(z)$ produced are `adequately shaped': the shape of the PIT distribution can tell us if the $p(z)$ produced are generally too wide/narrow, or if the $p(z)$ are over/under-predicting the true redshift.

The CRPS on the other hand tells us how well the $p(z)$ encapsulates or predicts the true redshift ($z_\rf{spec}$). The CRPS of a $p(z)$ can be expressed as
\begin{equation} 
\rf{CRPS} = \int^\infty_{-\infty} \left| C(z)- \mathcal{H}(z-z_\rf{spec}) \right|^2\, dz\ ,
\end{equation}
where $\mathcal{H}(z-z_\rf{spec})$ is the Heaviside step function with
\begin{equation} 
\mathcal{H}(z-z_\rf{spec}) = \left\{
\begin{array}{ll}
1, & z=z_\rf{spec} \\
0, & \rf{otherwise}.
\end{array}
\right.
\end{equation}
In this work, we use the symbol $\rho_\rf{CRPS}$ to represent the average CRPS value of all galaxies in the testing sample, in which the smaller the value, the better the $p(z)$ are at predicting their true redshifts. We refer the reader to \citet{polsterer_uncertain_2016} for a detailed description of both PIT and CRPS.

Finally, we also assess the quality of the redshift distribution $n(z)$. We can find how similar the \specz distribution $n_\rf{spec}(z)$ is compared to the \photoz distribution $n_\rf{phot}(z)$ by estimating $n_\rf{RMS}$, the root-mean-square difference between the distributions:
\begin{equation}
    n_\mathrm{RMS}=\sqrt{\int \left[ n_\mathrm{phot}(z)-n_\mathrm{spec}(z) \right]^2\, dz}.
\end{equation}
$n_\rf{RMS}$ provides us a quantitative measure to compare the performances of \photoz with distributions produced by different codes.

\subsection{Performance of Delight}\label{sec:quality}

\begin{figure*} 
\centering
\includegraphics[width=\linewidth]{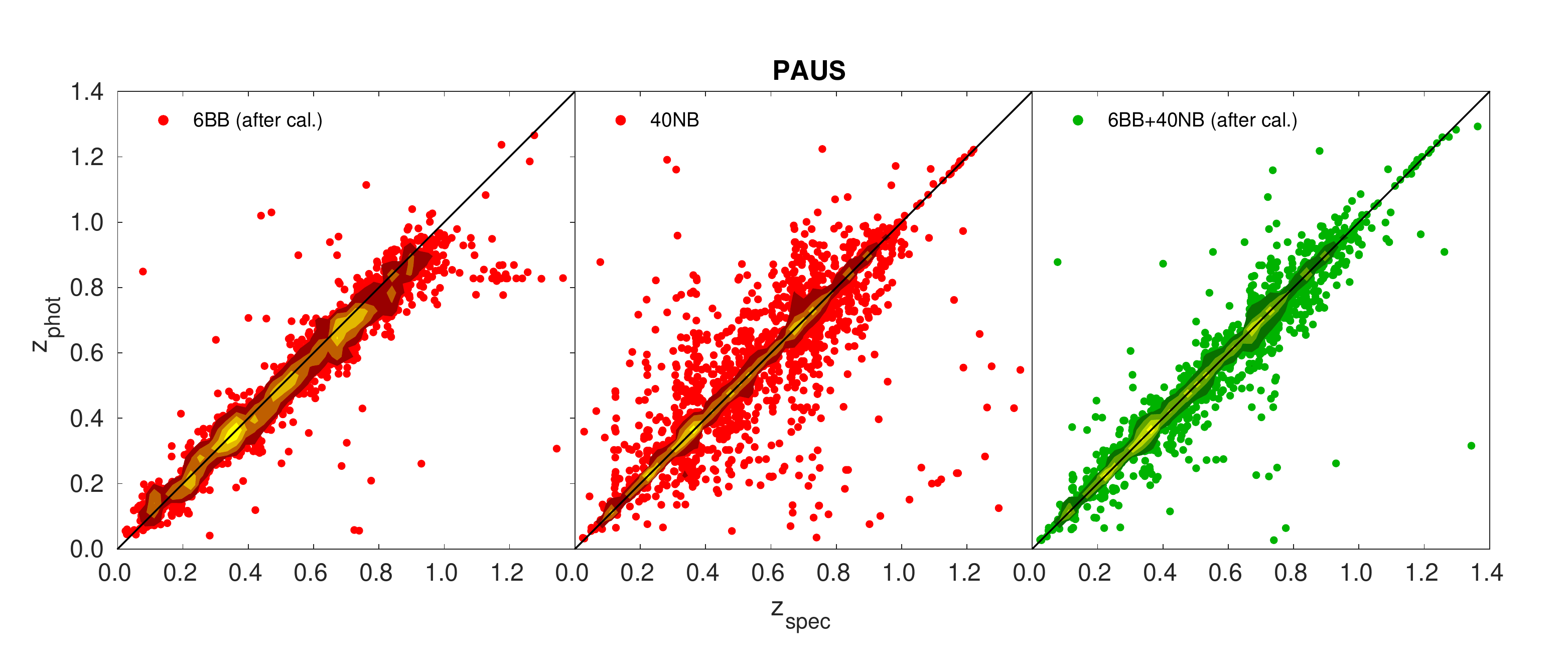}
\caption{Plot of \photoz vs. \specz, comparing the \photoz's when trained and tested using only $6$ $uBVriz$ broad bands (BB, left), only $40$ narrow bands (NB middle), and all $46$ bands combined (right). The flux calibration method was used for this plot.} \label{fig:res_delban}
\end{figure*}

Rows $1$ and $2$ from Table~\ref{tab:results} shows the \photoz's produced when only trained using the broad and narrow bands individually, and we find that by combining both broad and narrow bands (rows $4$ to $6$), we have achieved at least $34$ and $20$ per cent improvement in the \photoz scatter and $\sigma_{68}$, respectively (visualised in Fig.~\ref{fig:res_delban}). 

Rows $3$ to $7$ proceed to show the metrics for each calibration method, and on average, the performance of each method is quite similar, all within $4$ to $16$ per cent difference in $\sigma_\rf{RMS}$ and $\sigma_{68}$, respectively. Statistically, the flux calibration method seems to perform slightly better compared to the remaining ones, with the exception of the \photoz calibration method having better values of $\eta_\rf{out}$ and $n_\rf{RMS}$. This suggests that while the \photoz's produced by training with only $40$ narrow bands are not as competitive as when trained with all $46$ bands and calibrated broad bands (see Table~\ref{tab:results} and Fig.~\ref{fig:res_delban}), it is however sufficient to guide the calibration process. Note that we have also included the results of \delight run as a pure template code when calibrated using the flux calibration method for comparison, and we see that without the help of the GP, the \photoz results are similar for most metrics except a degradation in scatter of up to $-33.5$ per cent. Therefore the good results of \delight shown here are mainly due to the use of the Brown templates, the flux calibration, the combination of broad and narrow bands, and also the work of the GP.

As the three calibration methods presented in Section~\ref{sec:calibration} all result in very similar \photoz performance, we will only show results for the flux calibration method in the following. It is notable however that in all cases, the \photoz requirement of $\sigma_{68}<0.0035(1+z)$ is achievable for all objects at $i_\rf{auto}<20.0$, or objects with a $40$ per cent $\Theta$ cut at $i_\rf{auto}<22.5$. All three methods also shows that despite such high percentage $\Theta$ cuts being implemented, a significant number of high \photoz objects still remain in the sample.

\subsection{Comparison with Other Algorithms}\label{sec:comparison}

\begin{figure*} 
\centering
\includegraphics[width=\linewidth]{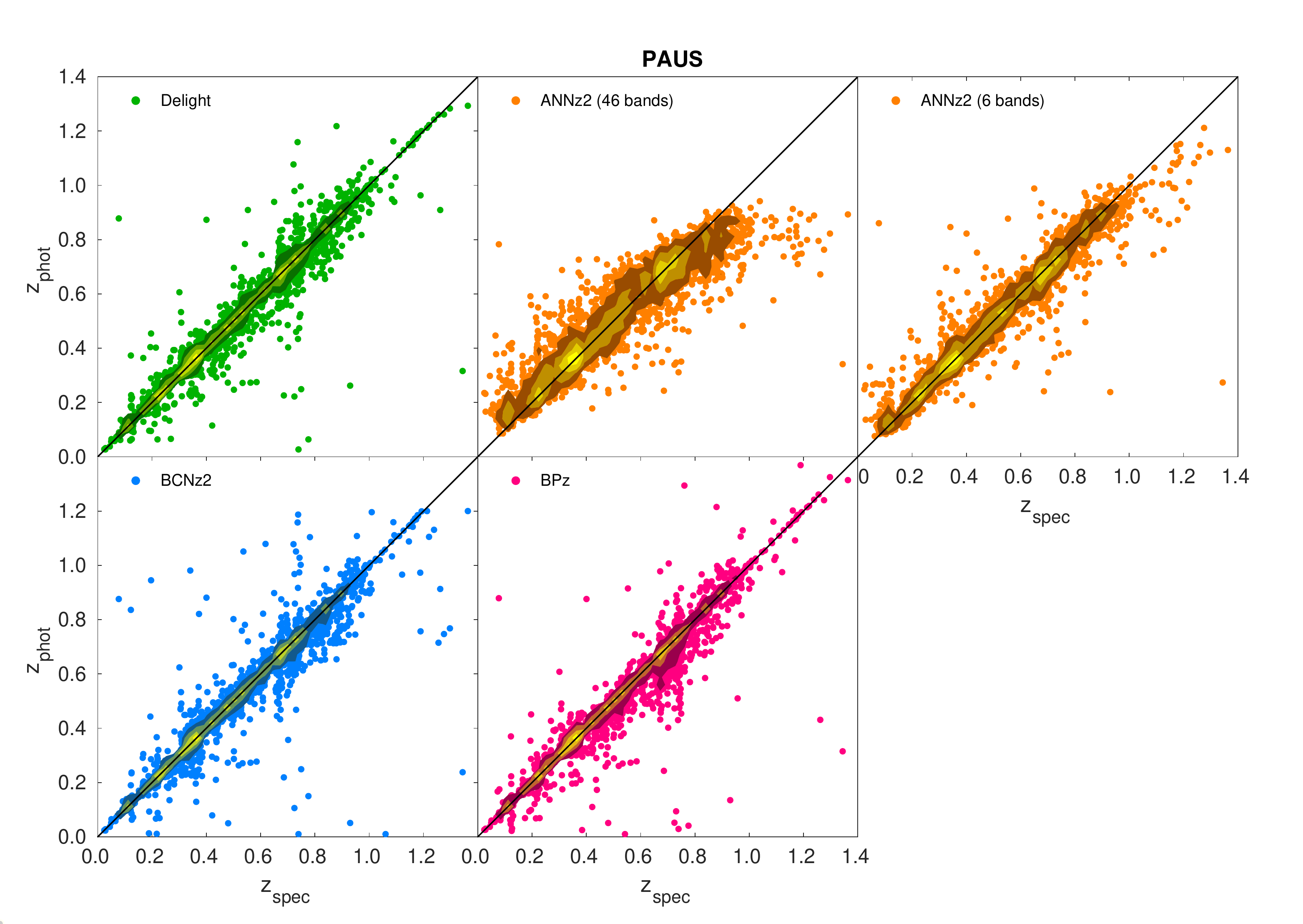}
\caption{Plots of \photoz vs. \specz, comparing the results of the flux calibration method of \delight (green), \annztwo trained with $46$ bands / $6$ broad bands (orange), \bcnz (blue) and \bpz (magenta). The same colouring scheme will be used to represent the respective methods in the following plots.} \label{fig:res_photoz}
\end{figure*}

Since the \delight results for each of the three calibration methods are very similar to each other, we decided to select only the flux calibration method to be compared to the results obtained by the two other algorithms used in this work, \annztwo and \bpz. We also include the point estimates from \citet{eriksen_pau_2019}. The values of $\sigma_\rf{RMS}$, $\sigma_{68}$ and other relevant metrics obtained from these algorithms are shown in rows $8$ to $11$ of Table~\ref{tab:results}, and visualised in Fig.~\ref{fig:res_photoz}.

From the figure, it is found that \annztwo, being a purely machine-learning based algorithm, is underperforming compared to the other algorithms. This machine-learning method is unable to make full use of the extra information provided by the $40$ narrow bands, and is shown to perform better without them. This is partially due to the problem of the curse of dimensionality \citep{bellman_dynamic_1957}, sharply diluting the pattern recognition power of the algorithm as the number of inputs increases. Besides, the very small training sample size may have heavily affected the potential of \annztwo. Here we note however that the deep learning code \textsc{deepz} is shown to work well on a similar sample \citep{eriksen_pau_2020}, therefore we hope to do follow-up evaluations of \annztwo on PAUS data in the future when a larger training set is available.

\begin{figure*} 
\centering
\includegraphics[width=\linewidth]{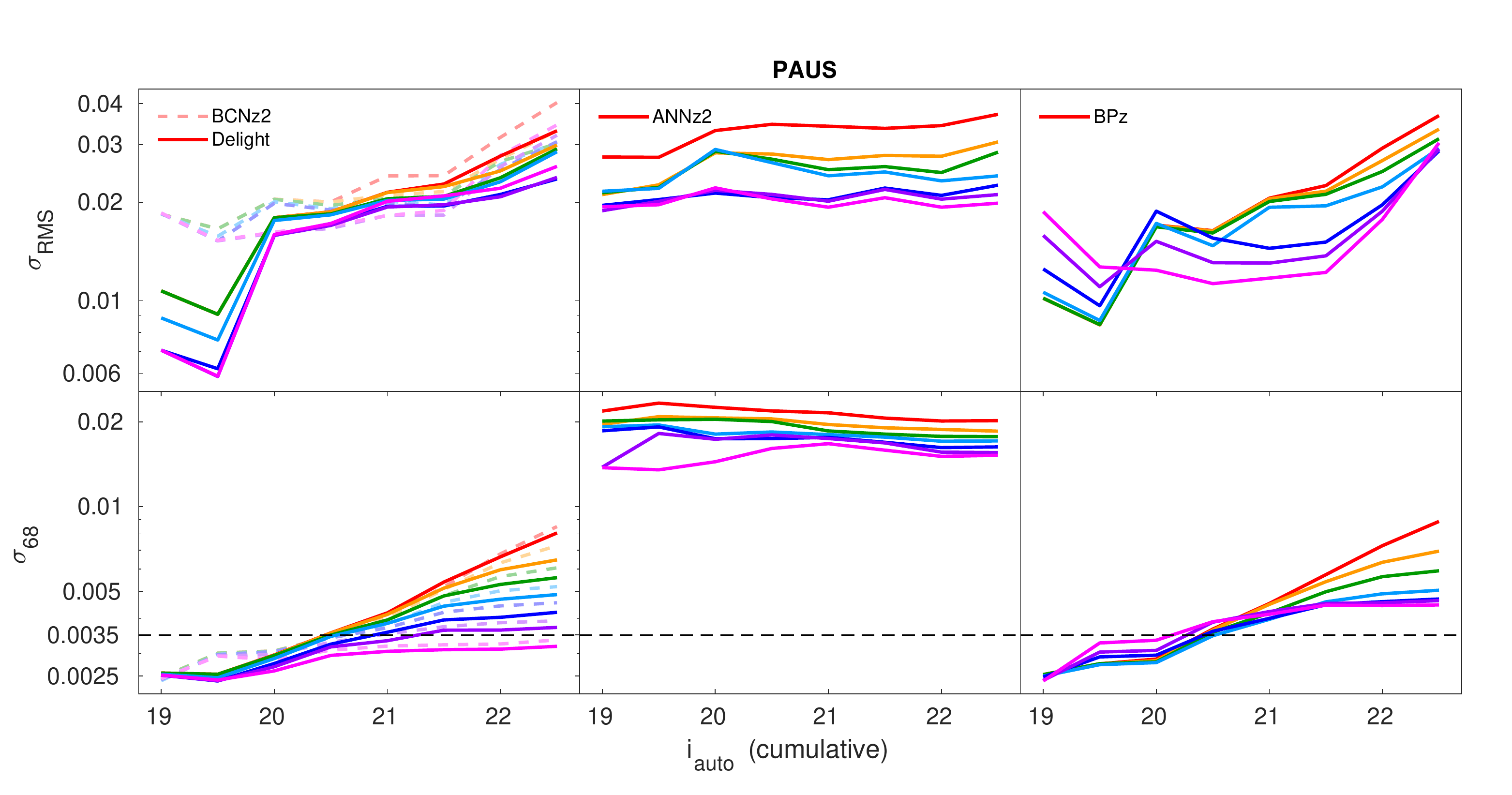}
\caption{Plots of $\sigma_\rf{RMS}$ (top) and $\sigma_{68}$ (bottom) with respect to $i_\rf{auto}$ (cumulatively), comparing the performance of \delight (left) with \annztwo (middle), \bpz (right) and \bcnz (left, dashed lines). The coloured lines represent the sample when cut systematically in the Bayesian odds ($\Theta$), keeping only objects with the best $100\%$ (red), $90\%$ (orange), $80\%$ (green), $70\%$ (blue), $60\%$ (navy), $50\%$ (purple) and $40\%$ (magenta) values. The black horizontal dashed line with $\sigma_{68}=0.0035(1+z)$ represents the \photoz quality target of PAUS for $50$ per cent of the objects at $i\sim 22.5$.} 
\label{fig:res_meti2}
\end{figure*}

\begin{figure*} 
\centering
\includegraphics[width=0.71\linewidth]{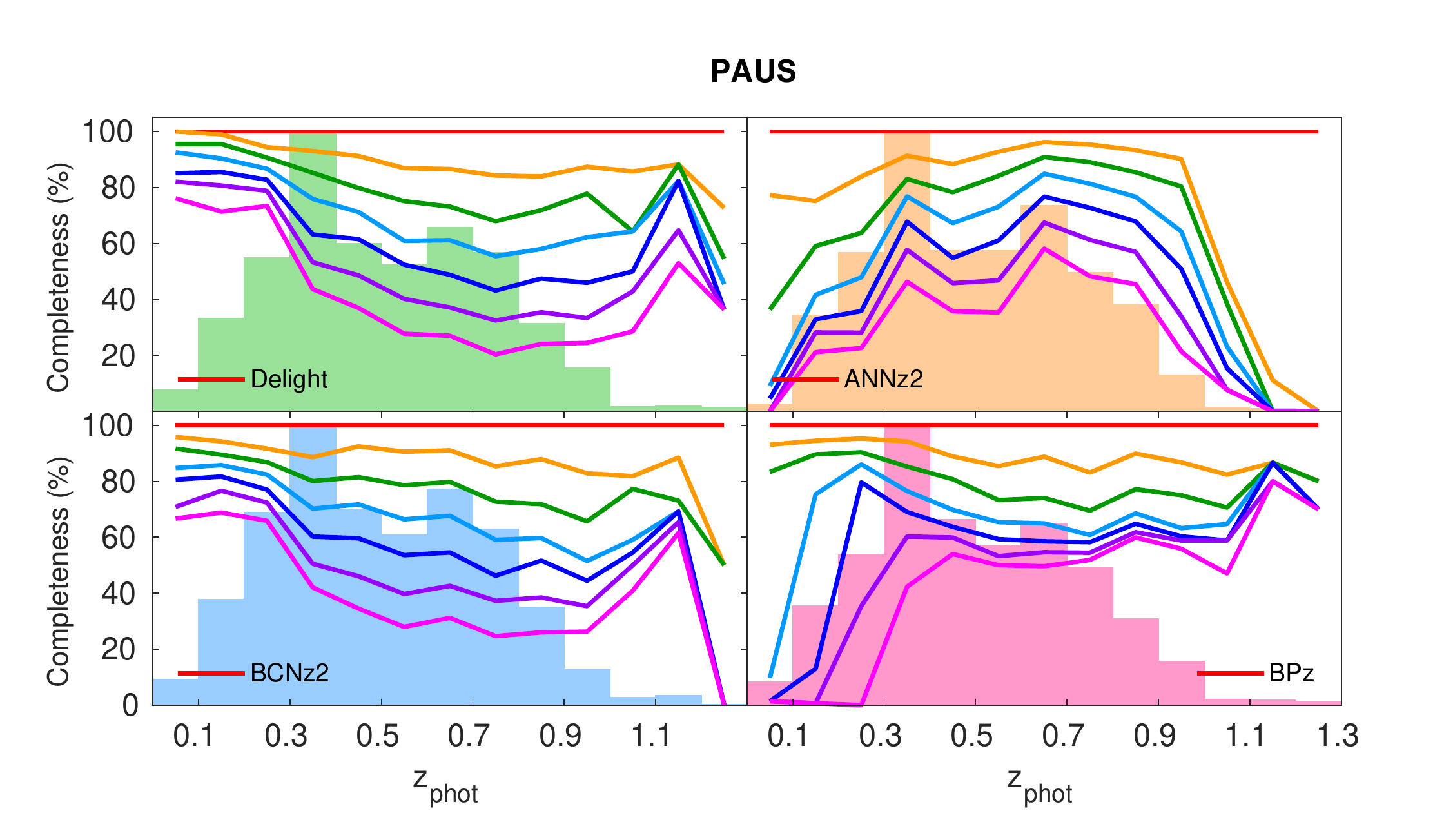}
\caption{Plot of percentage of objects within each \photoz bin with respect to the cut in $\Theta$ value for the results of \delight (top left), \annztwo (top right), \bcnz (bottom left) and \bpz (bottom right). The lines use the same colour scheme as those in Fig.~\ref{fig:res_meti2}, while the histograms in the background show the \photoz distribution for each method (relative number of objects in each \photoz bin).} \label{fig:res_metzp2}
\end{figure*}

In terms of the quality of the point estimate \photoz's, \delight is shown to fare well against \bcnz and \bpz (Fig.~\ref{fig:res_photoz}), both of which are purely template-based methods. As both \delight and \bpz used the same template sets in this case (i.e. the Brown templates), we find that the Gaussian process contributed to $25$ and $9$ per cent improvement in the scatter and $\sigma_{68}$, respectively, as compared to the pure template fit of \bpz. 

Despite the similarities in the point estimates for the entire sample (Table~\ref{tab:results}), when we cut the sample in percentages of $\Theta$ (Figs.~\ref{fig:res_meti2} and \ref{fig:res_metzp2}), we see two major differences. Firstly, the cut in $\Theta$ for \bpz does not systematically remove objects with high uncertainties (especially for objects brighter than $i_\rf{auto}=21$); and secondly, the cut in $\Theta$ for \bpz selectively removes objects with lower \photoz. In both cases, \delight is shown to not only perform better in this regard as compared to \bpz, but also better than all other algorithms shown.

\begin{figure} 
\centering
\includegraphics[width=0.99\linewidth]{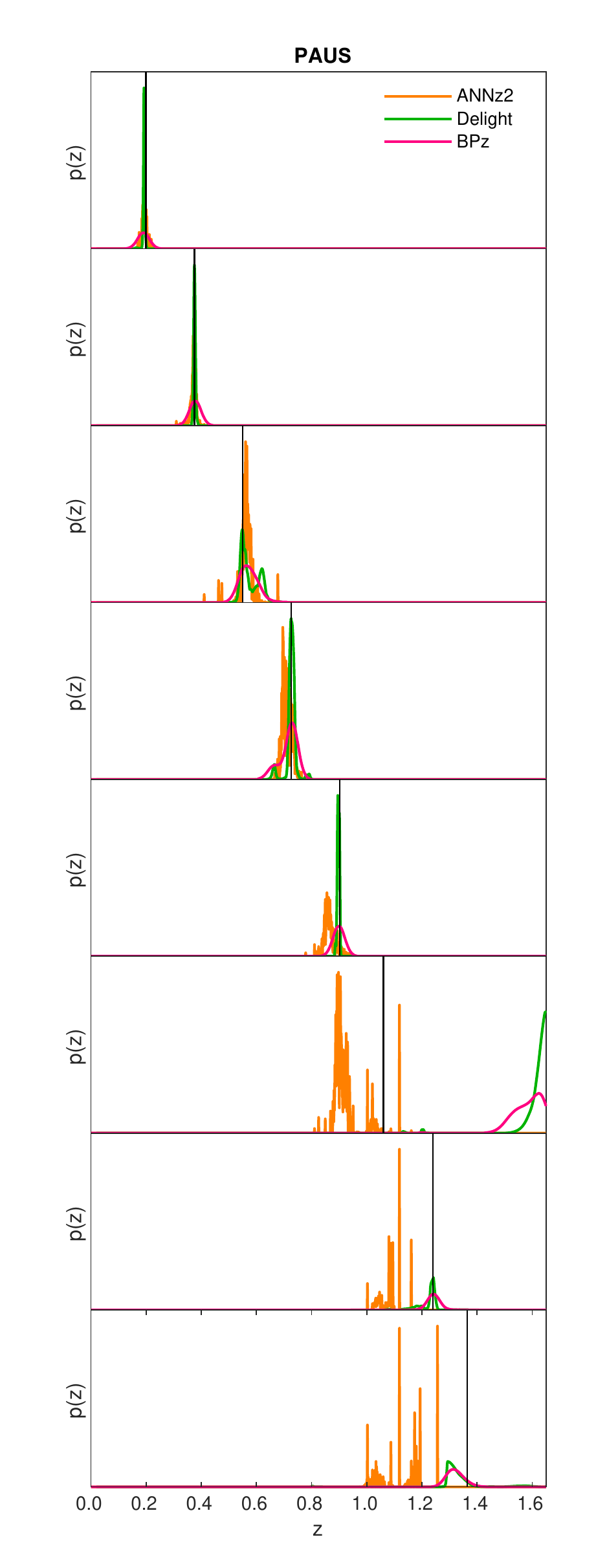}
\caption{Sample redshift PDF $p(z)$ for the \annztwo (orange), \delight (green) and \bpz (magenta). The black vertical lines shows the positions of the spectroscopic redshifts.} \label{fig:res_pz2}
\end{figure}

\begin{figure*} 
\centering
\includegraphics[width=\linewidth]{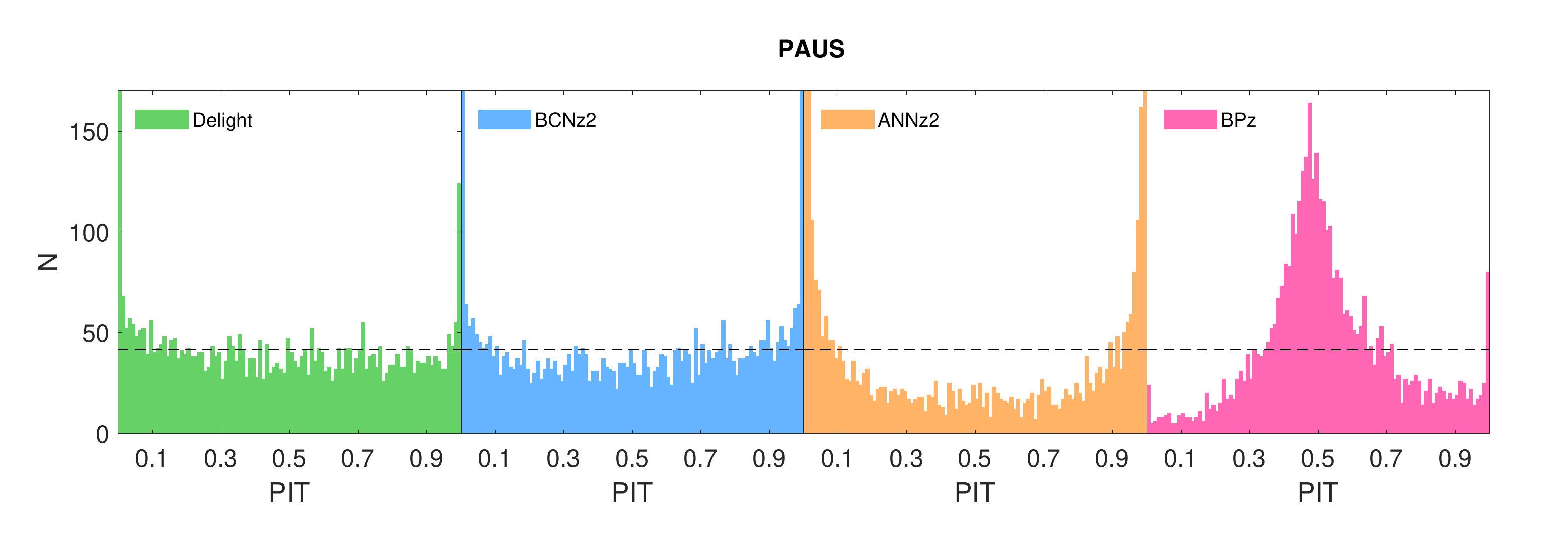}
\caption{Probability integral transform (PIT) distributions for the $p(z)$ produced by the four different algorithms, \delight (green), \bcnz (blue), \annztwo (orange) and \bpz (magenta). The dashed horizontal line indicates the mean of the distribution, and a flat distribution is ideal. A U-shaped distribution indicates that the $p(z)$ produced are too narrow, while a mountain-peak shaped distribution indicates that the $p(z)$ produced are too wide.} \label{fig:res_pit}
\end{figure*}

A selection of sample $p(z)$ produced by each algorithm is shown in Fig.~\ref{fig:res_pz2}, while the overall quality of the $p(z)$ produced are visualised in the PIT plots as shown in Fig.~\ref{fig:res_pit}. Once again we see \delight on average producing superior $p(z)$ compared to \annztwo and \bpz: it is obvious from the PIT plots that the $p(z)$ produced by \annztwo are too narrow (a U-shaped distribution), while those by \bpz are too wide (a significant central peak). In terms of $\rho_\rf{CRPS}$ (see Table~\ref{tab:results}), \delight once again performs better than both \bpz and \annztwo, where the adequate shapes and accurately positioned peaks of the $p(z)$ provide good predictions of the true redshift. 

We note that the $p(z)$ produced by \annztwo are ragged compared to \bpz and \delight, this is due to the limited training sample size and the low number of network committees used. We intend to look into several methodologies to smoothen machine-learning based $p(z)$ which are limited by such conditions; this is left for future work. The limited testing size has also produced an $n_\rf{spec}(z)$ distribution which is not smooth, thus despite \annztwo producing an $n(z)$ closest to the spectroscopic distribution (lowest $n_\rf{RMS}$), it may have experienced overfitting. Having said that, for the different \delight runs shown in Table~\ref{tab:results}, the values of $n_\rf{RMS}$ are consistent with the other metrics. Therefore, we leave the analysis of $n(z)$ to future work when a large enough testing sample is available.

\section{Application: identifying photo-z outliers}\label{sec:app}

\subsection{Analysing the photo-z outliers of Delight and BCNz2}\label{sec:outlier}
As we compared the \photoz results, we discovered that there are some galaxies that have similar \delight and \bcnz \photoz values, however these redshift values are far from their respective zCOSMOS spectroscopic redshifts or broadband \photoz's. Since both \bcnz and \delight utilise the PAUS narrow bands, we expect that the \photoz's they produce are more sensitive to emission lines as compared to \photoz's produced using only broad bands. Therefore, we suspect that objects that have similar \photoz values for \delight and \bcnz but have disagreeing \specz values to be an indication of either having (1) a catastrophic zCOSMOS spectroscopic redshift\footnote{While we have already selected to use only secure spectroscopic redshifts in this work, we still deem this as a possibility, since a $1$ per cent outlier rate in $4000+$ \specz measurements may still yield $40$ objects, which is within the same order of number of objects being investigated in this section. Our results later in this section however have verified that most of the outliers are not caused by catastrophic spectroscopic redshifts.}, (2) outlier broadband or narrowband fluxes, or (3) misidentification of close neighbours.

For the purpose of this inquiry, we have selected $30$ objects from the sample which are \photoz outliers in $z_\rf{Delight}$ vs. $z_\rf{spec}$ or $z_\rf{BCNz}$ vs. $z_\rf{spec}$, yet
are not outliers in $z_\rf{Delight}$ vs. $z_\rf{BCNz}$. Mathematically, they satisfy the following conditions:
\begin{enumerate}
\item $\frac{\left| z_\rf{Delight}-z_\rf{spec}\right|}{1+z_\rf{spec}}\ge 0.15$ or $\frac{\left|z_\rf{BCNz}-z_\rf{spec}\right|}{1+z_\rf{spec}}\ge 0.15$, and
\item $\frac{\left| z_\rf{Delight}-z_\rf{BCNz}\right|}{1+\frac{z_\rf{Delight}+z_\rf{BCNz}}{2}} < 0.15$.
\end{enumerate}
Note that the $z_\rf{Delight}$ used here refers to the \photoz produced using the flux calibration method, trained using $46$ bands guided by the Brown templates.

\begin{figure*} 
\centering
\includegraphics[width=\linewidth]{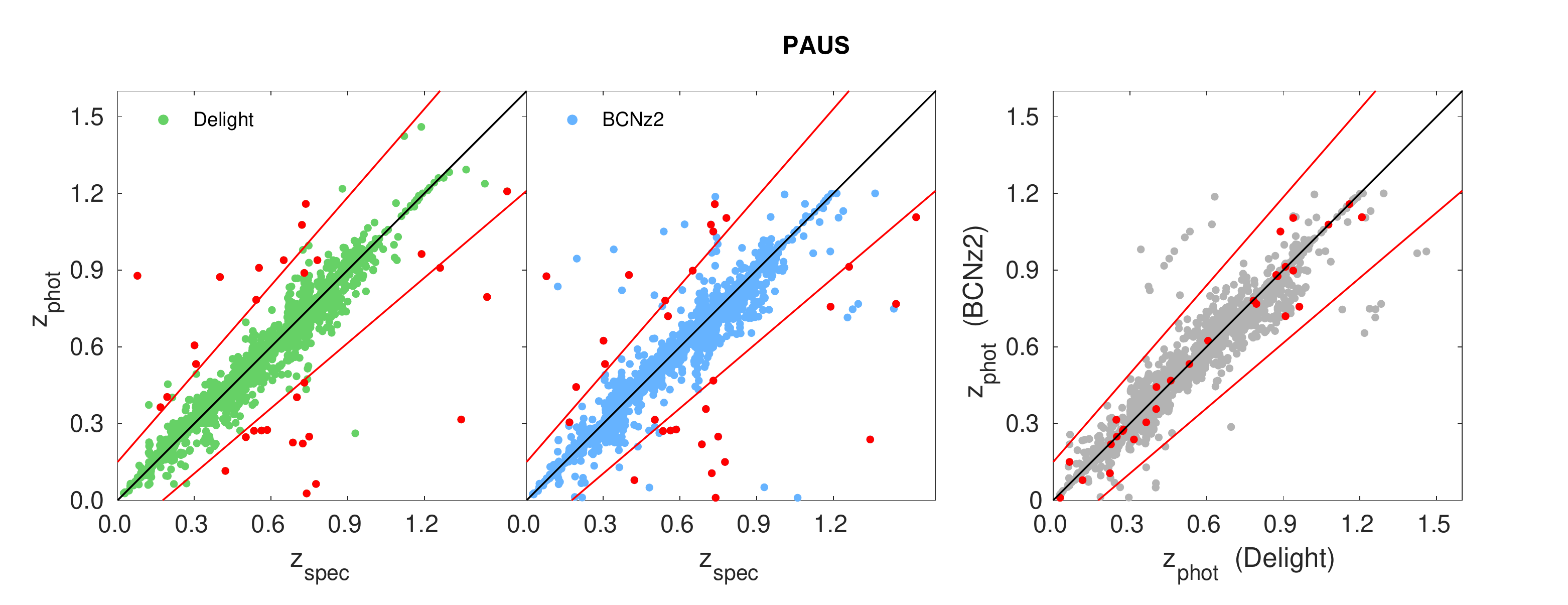}
\caption{The selected $30$ objects (red dots) marked for this outlier analysis. These objects are \photoz outliers of either \delight or \bcnz with respect to $z_\rf{spec}$, but are not outliers with respect to each other.} \label{fig:outliers1}
\end{figure*}

These $30$ objects are visualised in the redshift-redshift plots in Fig.~\ref{fig:outliers1}. Note that in the following paragraphs, we will define a \photoz to be \textit{catastrophic} if it is found to be an outlier with respect to its \specz, as defined mathematically above. These objects are found to have faint magnitudes ($i_\rf{auto}>19.75$) and small angular sizes ($r_{50}<60$ ACS pixels, or $1.8''$), which describe most galaxies of interest for PAUS. We study several different attributes of these objects, namely their respective \photoz's by \delight, \bcnz and \lephare, \photoz PDFs, best-fit templates (Brown and GP), spectra and images. We summarise important observations according to their respective attributes below.

\textit{\Photoz's}. While these $30$ objects have been identified as outliers when trained using $46$ bands, we find that two-thirds of these objects have non-catastrophic \photoz's when trained with either only the broad or narrow bands, respectively. In other words, only one-third of these objects have catastrophic \photoz's regardless of which bands were used in the training or fitting process. This suggests that most of the time, outlier fluxes in the broad or narrow bands may have caused a degradation in \photoz quality when trained together (more on this in the \textit{templates} paragraph below). We have also made a comparison between \delight \photoz's with those produced by \lephare for the COSMOS2015 catalogue \citep{laigle_cosmos2015_2016}, and found that in fact half of the $30$ objects have non-catastrophic \lephare \photoz's. This suggests that the infrared $yJHK$ bands could have played a role in improving the PAUS \photoz's, and could be incorporated in future trainings in case the PAUS photometry is problematic\footnote{We note that these additional bands will not be available over most of PAUS, which targets Canada-France-Hawaii Telescope Legacy Survey (CFHTLS) wide fields W1 to W4. There is however some infrared data on these fields provided by the Wide-field InfraRed Camera (WIRCam) and the VISTA Kilo-degree Infrared Galaxy Survey (VIKING).}.

\textit{\Photoz PDFs}. We inspected the secondary/tertiary peaks of the PDFs for all \delight runs (trained with $6$ broad bands, $40$ narrow bands, or both), and find that less than $20$ per cent of these secondary/tertiary peaks coincide with their respective \specz's. We deduce that despite the importance of secondary PDF peaks in redshift distributions, they do not significantly influence the \photoz quality of these $30$ objects.

\textit{Templates}. \delight utilises the $129$ \citet{brown_atlas_2014} templates and the $4203$ training objects to guide the GP to produce the same number of new flux-redshift templates, which are used to produce \photoz's for the objects. In the training process, \delight would always choose one best-fit Brown template for each training galaxy to be trained by the GP. Here we inspected two different kinds of best-fit Brown templates to these $30$ outliers: one fixed at the \specz, and the other with the redshift as a free parameter. In both cases, we examined
\begin{enumerate}
    \item if the objects fit to the same templates when trained with only broad bands, only narrow bands, or both, respectively;
    \item if there are any trends in galaxy morphological types, based on the galaxy type classification indicated by the template;
    \item if there is any correlation between the $\chi^2$ value of the best-fit templates and the quality of \photoz's; and
    \item if any outlier narrowband fluxes can be identified as the cause of the degradation of \photoz.
\end{enumerate}

As expected, we find that $70$ per cent of the outlier objects have different best-fit Brown templates between the fits at fixed \photoz and \specz, which contrasts with the case for non-outliers at only $35$ per cent. We also find that only slightly more than a third of both the outlier and non-outlier objects were fitted to the same templates when trained using broad bands as compared to trained with all $46$ bands. The high percentage of objects with different template fits at different reference redshifts (\photoz or \specz) and flux combinations (broad bands, narrow bands, or both) also resulted in no trend in galaxy morphological types among the outliers. 

However, it was found that up to $60$ per cent of the objects have their best-fit template $\chi^2$ value correlating with the quality in \photoz, which further affirms the usage of this as a metric to remove unreliable \photoz's (see Section~\ref{sec:newmetrics}), as also attempted by \citet{eriksen_pau_2019} and \citet{eriksen_pau_2020}.

\begin{figure} 
\centering
\includegraphics[width=\linewidth]{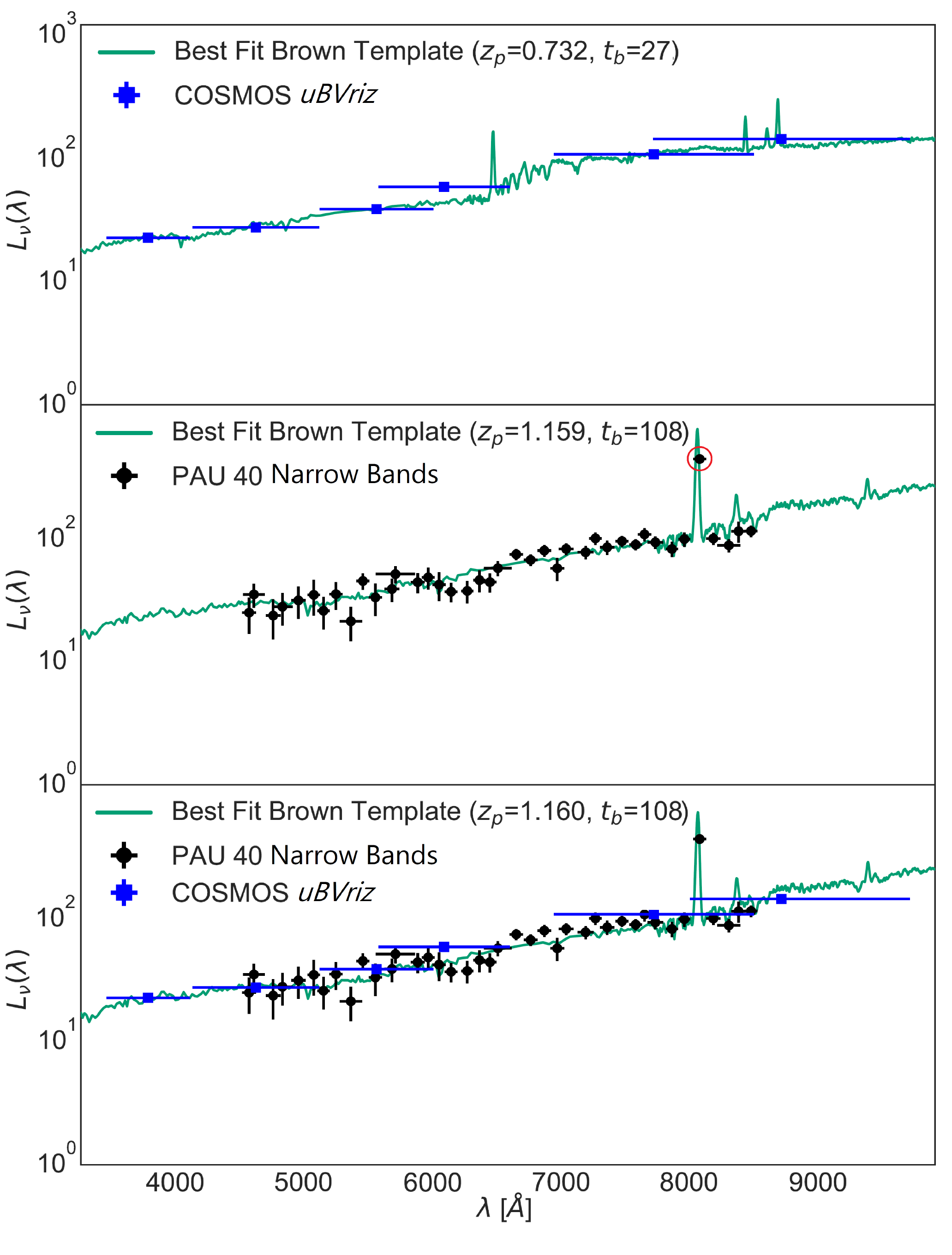}
\caption{A sample of best-fit Brown templates (unfixed redshift) when fit to only broadband fluxes (top), only narrowbands fluxes (middle), and both fluxes (bottom) for the galaxy with zCOSMOS ID $805216$. $L_{v}(\lambda)$ is the rest-frame luminosity density (or SED) of the galaxy. This galaxy has $z_\rf{spec}=0.736$, $z_\rf{p}$ and $t_\rf{b}$ in the figure refer to its \photoz and best-fit Brown template number, respectively. The outlier narrowband flux shown in the middle panel (red circle) has caused a misfit in template type, resulting in erroneous \photoz's for both cases.} \label{fig:tempfit}
\end{figure}

Perhaps a more significant finding from the study of the best-fit templates is the ability to identify outlier narrowband fluxes. Fig.~\ref{fig:tempfit} shows an example which highlights the importance of identifying outlier narrowband fluxes, which is shown to significantly affect the \photoz results. It was found that a third of the $30$ objects contained outlier narrowband fluxes, which results in entirely different template fits and \photoz's when trained with narrow bands, as compared to when trained with broad bands only. Among these $10$ objects, $8$ of them are shown to have worse \photoz as compared to training without the narrow bands. We find indications for a significant fraction of narrowband flux outliers also for galaxies without catastrophic redshift failures. Forthcoming PAUS data reductions will therefore implement methods to identify and correct flux outliers.

\textit{Images}. We inspect the individual object images compiled by zCOSMOS DR$3$, these are $5''\times 5''$ images observed by the Hubble Space Telescope/Advanced Camera for Surveys (HST/ACS) in the F$814$W filter \citep{koekemoer_cosmos_2007}. Among the $30$ outlier objects, we find $63.3$ and $26.7$ per cent of them having bright neighbours within $5''$ and $3''$ of the primary source, respectively. Having said that, we have not found any correlation between the presence of bright neighbours to the other attributes that we have studied thus far. In fact the opposite is true: we find that $60$ per cent of the objects with outlier narrowband fluxes actually have primary sources without any bright neighbours in vicinity.

\begin{figure*} 
\centering
\includegraphics[width=0.9\linewidth]{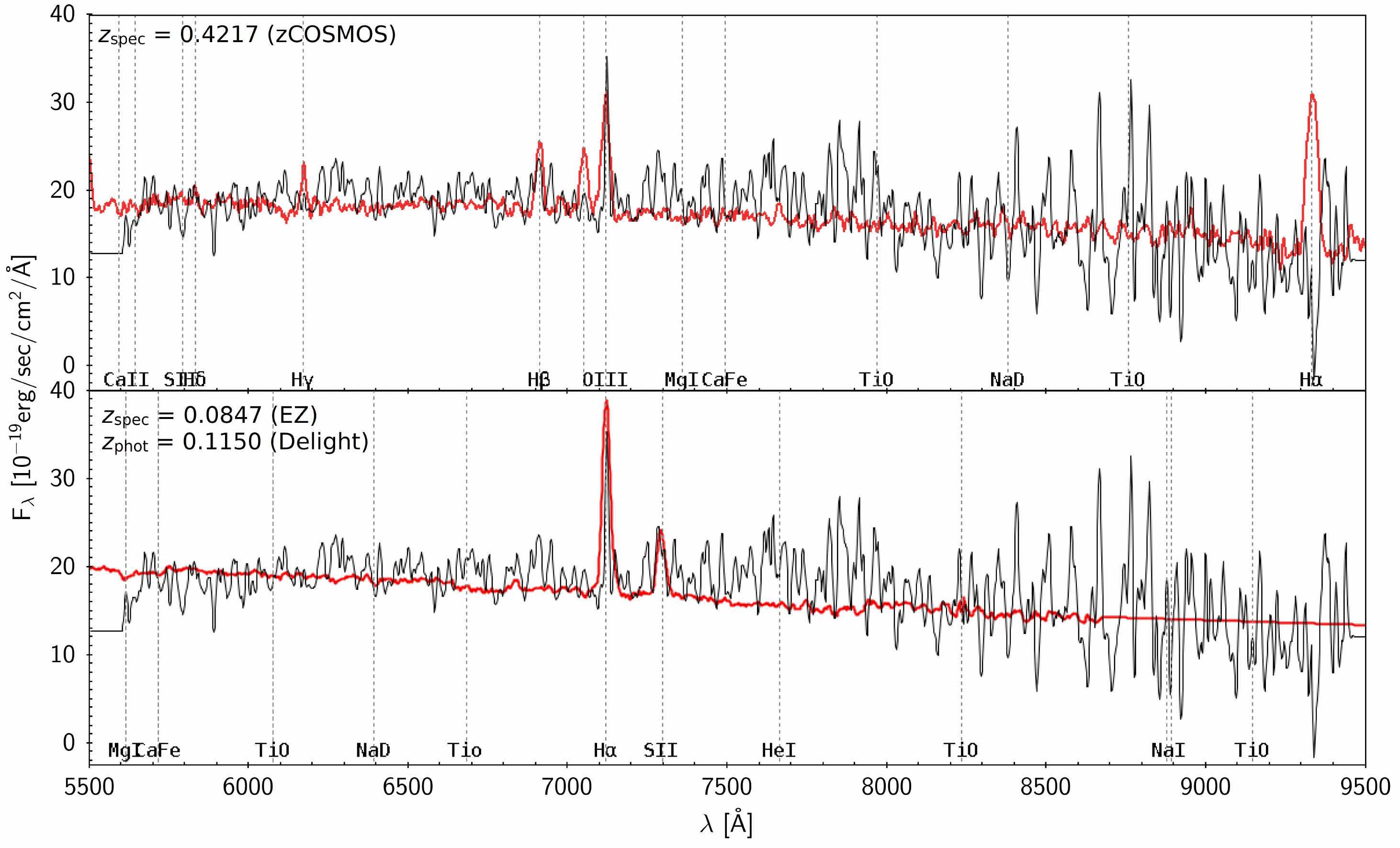}
\caption{Spectral line fitting (red) for the original spectra (black) of the galaxy with zCOSMOS ID $804179$. The \specz given by zCOSMOS is $0.4217$ (top), while the best-fit using \textsc{ez} \citep{garilli_ez_2010} gives a \specz of $0.0847$ (bottom), which is closer to the \photoz value of $0.1150$ estimated by \delight.} \label{fig:ezfit}
\end{figure*}

\textit{Spectra}. So far we have assumed that the zCOSMOS spectra obtained are reliable, as only entries with high-confidence quality flags have been selected for training (see Section~\ref{sec:zcosmos}). In order to probe further, we examined the one-dimensional spectra obtained by the VIMOS spectrograph, which is processed by the VIMOS Interactive Pipeline and Graphical Interface \citep[VIPGI,][]{scodeggio_vvds_2005} to produce the zCOSMOS \specz's used in this work. The spectra have a range between $5500$~\r{A} and $9450$~\r{A}, measured with a resolution of $R\sim 600$ at $2.5$~\r{A} per pixel \citep{lilly_zcosmos_2009}. 

We used the redshift measurement tool \textsc{ez} \citep{garilli_ez_2010} to inspect the spectra of the $30$ outlier objects, and compared our best fits to the spectroscopic redshift produced by zCOSMOS, and also the \photoz's produced by \delight, \bcnz, \deepz, \lephare (COSMOS2015) and those of \citet{alarcon_pau_2021}. 

Upon inspection, we find that up to $10$ of these objects ($33$ per cent) have disputable zCOSMOS \specz (e.g. two possible redshift values, different best-fit redshift values, line confusion, and low signal-to-noise). However, most of these potential \specz failures could be forced-fitted to the zCOSMOS \specz and still look satisfactory, which leaves only $2$ ($6.7$ per cent) of these objects having truly catastrophic \specz's. Both these objects are found to have better \textsc{ez} fits at redshift values within $10$ per cent uncertainty from the \photoz's produced by \delight and other algorithms. The spectrum of one of these objects is shown in Fig.~\ref{fig:ezfit}. We have also found one isolated case where the spectra belonged to a bright neighbour and has been mismatched to the PAUS photometry.

Generally, the higher-redshift objects are identified by clear \textsc{O II} ($3727.1$~\r{A}) emission lines, while the lower-redshift objects are identified by clear H$\alpha$ ($6564.6$~\r{A}) emission lines. We therefore conclude that although catastrophic \specz's played a role in this situation, our results did not provide enough evidence to say that it is a major cause for catastrophic \photoz's produced by \bcnz and \delight. This is not surprising since we have only selected secure spectroscopic redshifts from COSMOS to be used in this work. However this highlights the usefulness of multiple PAUS \photoz's being used to determine failure rates in insecure spectroscopic redshifts.

To summarise this part, we believe that the potentially important source for catastrophic \photoz's in the context of PAUS are the outlier narrowband fluxes, with weak evidence for the existence of a small number of \specz failures. We leave the tackling of outlier narrowband fluxes to future work, but in the following section, we attempt to improve our process to identify and remove these outlier \photoz's.

\subsection{New metrics to remove photo-z outliers} \label{sec:newmetrics}
In Figs.~\ref{fig:res_meti2} and \ref{fig:res_metzp2} we have used the Bayesian odds ($\Theta$) to cut the sample, and the aim of this was to keep as many objects as possible while achieving the goal of $\sigma_{68}\le0.0035(1+z)$. Here, we extend our previous results further towards that goal by introducing several new metrics to better separate the \photoz outliers from the sample. These metrics are motivated by the inspection of the $30$ outliers in Section~\ref{sec:outlier}, and they are defined as follows:

\begin{enumerate}
    \item The \textit{Delight-BCNz2 metric} ($\Delta_\rf{DB}$),
    \begin{equation}
        \Delta_\rf{DB} \equiv \frac{\left| z_\rf{Delight}-z_\rf{BCNz} \right|}{1+\frac{z_\rf{Delight}+z_\rf{BCNz}}{2}},
    \end{equation}
    a metric used to identify the similarity between \delight and \bcnz \photoz's. It is plausible that, in general, the closer the \photoz's between the two algorithms, the more reliable they are;
    
    \item The \textit{Delight \photoz standard deviation} ($\sigma_\rf{D}$), which is the standard deviation between all \delight \photoz runs regardless of calibration method and number of bands. Smaller deviations could indicate more reliable \photoz's;
    
    \item The \textit{chi-squared value of the best-fit Brown template} ($\chi^2_\rf{t}$), where we identified a trend that the better the fit, the more reliable the \photoz; and
    
    \item The \textit{broadband-narrowband complementary metric} ($\rho^2$),
    \begin{equation}
        \rho^2 \equiv \int p_{\rf{BB}}(z)p_{\rf{NB}}(z)\, dz,
    \end{equation}
    where $p_\rf{BB}(z)$ and $p_\rf{NB}(z)$ are the $p(z)$ produced by \delight when trained with only broad bands and only narrow bands, respectively. By multiplying these two $p(z)$ and summing over the distribution at each step $i$, we can identify the consistency between the broadband and narrowband $p(z)$. A higher value of $\rho^2$ means a larger overlap, which indicates more reliable \photoz's.
\end{enumerate}

Together with $\Theta$ and the \delight \photoz error ($\delta z$), we yield a total of $6$ metrics to experiment with. Using the results from the flux calibration method, we generate and test the individual performance for each of these metrics. For each metric, we measure the $\sigma_\rf{RMS}$ and $\sigma_{68}$ after systematically removing objects with the worst metric values, $10$ per cent of the total sample size each time, until we reach a sample size of only $40$ per cent. 

\begin{figure*} 
\centering
\includegraphics[width=\linewidth]{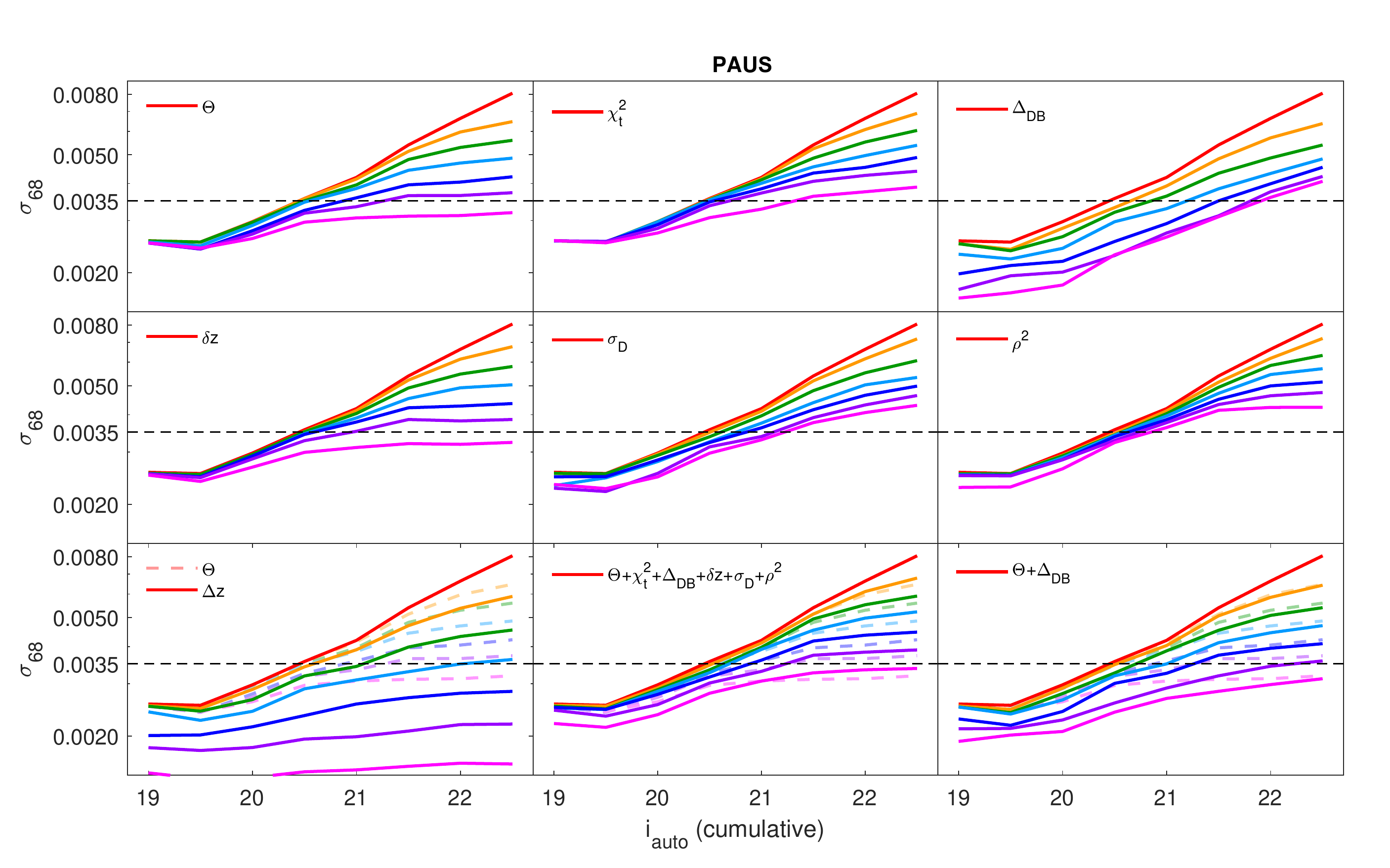}
\caption{Plot of $68$th percentile error ($\sigma_\rf{68}$) vs. $i_\rf{auto}$ (cumulative) when cut using the following metrics: the Bayesian odds ($\Theta$), best-fit Brown template $\chi^2_\rf{t}$ value, \delight-\bcnz metric ($\Delta_\rf{DB}$), \delight \photoz error ($\delta z$) and standard deviation ($\sigma_\rf{D}$), and the broadband-narrowband complementary metric ($\rho^2$). The coloured lines follow the same percentile cuts as shown in Fig.~\ref{fig:res_meti2}, with the dotted-coloured lines in the background of the bottom panels depicting the results of $\Theta$ for easier comparison. The bottom-left panel shows the cut in $\Delta z$ (defined in Section~\ref{sec:metrics}), the unsurpassable theoretical best used for reference. The bottom-middle panel shows the cuts when all the above metrics were combined, while the bottom-right shows the combination of metrics which yield the best results.} \label{fig:newmetric2}
\end{figure*}

We also repeat the exercise by using combined cuts on several metrics, testing all $57$ combinations of the $6$ metrics. We note that we do not combine the metrics by averaging or multiplying them, as it would have diluted the impact of the individual metrics. Instead, we rank the values for each metric individually (from best to worst), and remove objects rank by rank, starting with metric values lying in the worst rank. E.g. for the combination of metrics $\Theta+\Delta_\rf{DB}$, we first remove all objects which share the worst values of $\Theta$ and $\Delta_\rf{DB}$, then remove all objects sharing the second worst values of them, and so on, until we reach a required sample size percentile ($90$, $80$, etc), where we output the values of $\sigma_\rf{RMS}$ and $\sigma_{68}$. We visualise the performance of these metric cuts at several percentiles for $\sigma_{68}$ with respect to $i_\rf{auto}$ (cumulative) in Fig.~\ref{fig:newmetric2}.

We find that each performance metric cuts the sample differently: while metric cuts of $\sigma_\rf{D}$ and $\rho^2$ reduce the scatter ($\sigma_\rf{RMS}$) significantly, metric cuts of $\Theta$ and $\Delta_\rf{DB}$ reduce the $\sigma_{68}$ instead. The metric $\chi^2_\rf{t}$, however does not seem to bring any significant improvement to the results. We have also plotted a cut in $\Delta z=\frac{\left| z_\rf{phot}-z_\rf{spec} \right| }{1+z_\rf{spec}}$ (bottom-left panel in Fig.~\ref{fig:newmetric2}), which is the theoretical `best metric', providing an upper limit to be compared with the performance of each of the metrics. Here we noticed that even with the theoretical best metric, a cut of slightly lesser than $70$ per cent (blue line) on the sample is still necessary to fulfil the PAUS target of $\sigma_{68}<0.0035(1+z)$ (dotted line) for \delight. 

Therefore, we select the $60$ per cent cut (navy line, retaining $60$ per cent of galaxies) as a benchmark to assess the performance of these metrics, we do so by locating where this line cuts the dotted line (i.e., finding the maximum value of $i_\rf{auto}$ where the \photoz's achieves the PAUS target at $60$ per cent cut). From Fig.~\ref{fig:newmetric2}, it is clear that cutting in all $6$ metrics does not necessarily outperform the performance when cutting with only $\Theta$, so we searched for the best combination of metrics for $\sigma_\rf{RMS}$ and $\sigma_{68}$ separately.

For $\sigma_\rf{RMS}$, the best combination of metrics is $\Delta_\rf{DB}+\sigma_\rf{D}+\rho^2$, and this combination achieves $\sigma_\rf{RMS}<0.0035(1+z)$ at $i_\rf{auto}<19.27$ at $60$ per cent cut, a significant improvement to the case when only $\Theta$ was used, where it did not cut the line at all. For $\sigma_{68}$, the best combination of metrics is $\Theta+\Delta_\rf{DB}$ where it reached $\sigma_\rf{68}<0.0035(1+z)$ at $i_\rf{auto}<21.25$ at $60$ per cent cut, which is also a significant improvement as compared to $\Theta$ at $i_\rf{auto}<20.88$. Here we note that in fact using $\Delta_\rf{DB}$ alone, the target can be reached at a higher limit of $i_\rf{auto}<21.50$, which highlights the significance of a synergy between \delight and \bcnz in selecting a high quality \photoz sample.

\begin{figure*} 
\centering
\includegraphics[width=\linewidth]{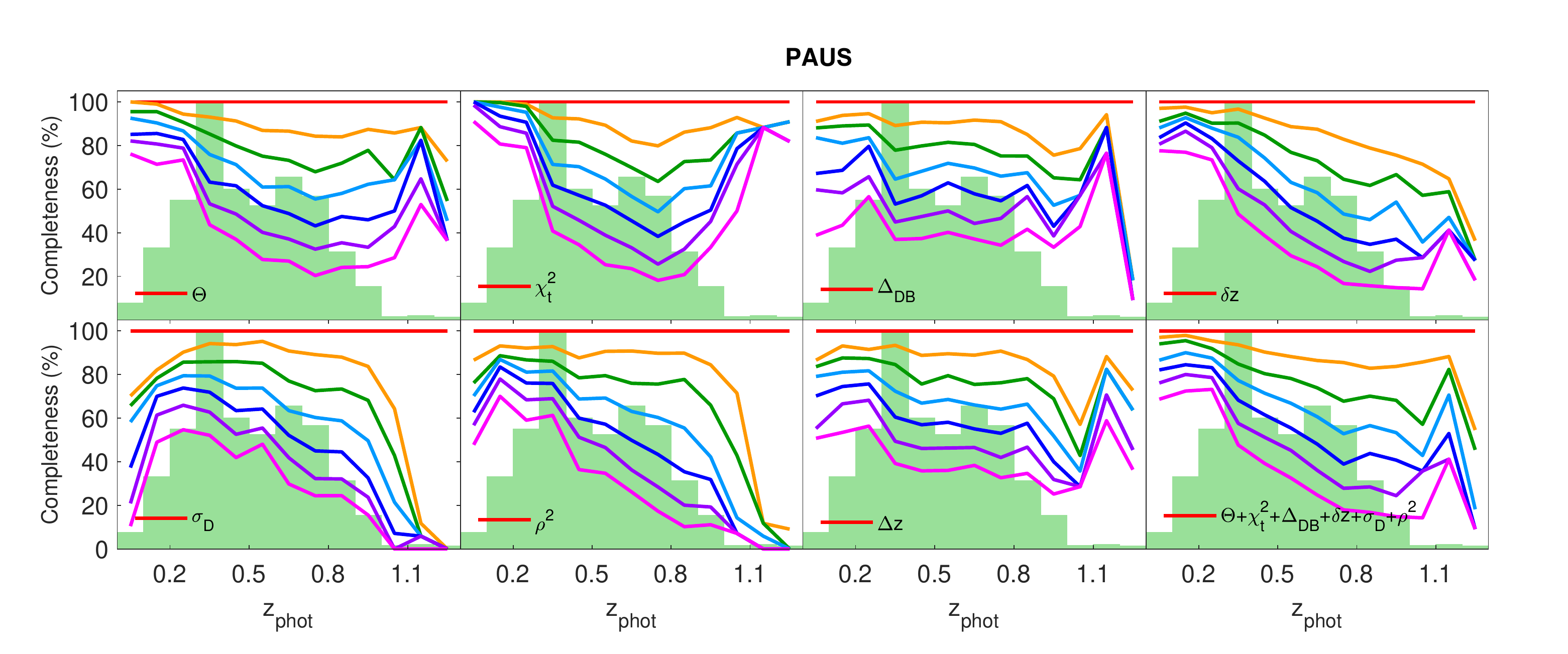}
\caption{Plot of percentage of objects within each \photoz bin with respect to the cut in performance metric values listed in Fig.~\ref{fig:newmetric2}. The lines show the percentiles of the same colour scheme as in Fig.~\ref{fig:res_meti2}, while the histograms in the background show the relative number of objects in each \photoz bin. The bottom-right plot shows when the combination of all $6$ metrics are used to cut the sample.} \label{fig:newmetric3}
\end{figure*}

Finally, we also show the performance of the metrics in terms of the completeness with respect to the \photoz (using \delight's flux calibration method), visualised in  Fig.~\ref{fig:newmetric3}. We find that metrics like $\sigma_\rf{D}$ and $\rho^2$ tend to selectively remove high \photoz objects, while $\Theta$, $\chi^2_\rf{t}$ and $\Delta_\rf{DB}$ tend to remove mid-ranged \photoz objects. In general, a cut using all $6$ performance metrics at $60$ per cent cut shows a balanced result in the completeness, keeping a sufficient number of high redshift objects in the sample.

To summarise the performance of the individual metrics, 
\begin{itemize}
    \item $\chi^2_\rf{t}$ is the least-performing metric here; it does not bring significant positive impact to the results;
    \item Cuts in $\sigma_\rf{D}$ and $\rho^2$ help to improve the scatter, however, they tend to selectively remove higher \photoz objects from the sample;
    \item $\Theta$ and $\delta z$ show very similar results, however $\Theta$ tends to keep more high \photoz objects in the sample; and
    \item $\Delta_\rf{DB}$ is the best-performing metric here, and we recommend the use of such a metric to remove outlier \photoz's from a sample.
\end{itemize}

\section{Conclusion and Future Work} \label{sec:conc}

In this work we have optimised \delight, a hybrid template-machine learning algorithm such that it could be used to obtain \photoz's for PAUS, by utilising its $40$ narrowband fluxes combined with $6$ $uBVriz$ COSMOS broadband fluxes. We have shown three distinct methods to calibrate the broadband and narrowband fluxes, and found that all three methods yield comparable results, although the most stable and the one which produces the lowest value of $\sigma_{68}$ is what we defined as the \textit{flux calibration method}: a method where we calibrate the broadband fluxes with respect to the narrowband fluxes by finding the flux ratio of the filter combinations which overlap. This calibration method is entirely photometric, and it was able to produce \photoz's with a scatter reaching as low as $\sigma_\rf{RMS}=0.0331(1+z)$ and $\sigma_{68}=0.0081(1+z)$  for the full PAUS galaxy sample at $i_\rf{auto}<22.5$.

We have also compared the results of \delight with a machine learning algorithm (\annztwo) and a template-based algorithm (\bpz and \bcnz). We find that \annztwo underperforms significantly, indicating that \annztwo in its basic form is not suitable for narrowband surveys with large number of bands and small number of training objects. 

Despite the \photoz performance of \bpz being within $9$ per cent difference of that of \delight, the latter still stood out in terms of the quality of the \photoz PDF $p(z)$ ($16$ per cent better in $\rho_\rf{CRPS}$) and the effectiveness of its Bayesian odds ($\Theta$) cut in retaining objects with higher quality \photoz without losing too many high-redshift objects. \delight is also shown to produce competitive results as compared to \bcnz ($5$ per cent lower in $\sigma_{68}$), the default \photoz produced for the PAUS.

Further investigation on the common \photoz outliers of \delight and \bcnz led to the conclusion that outlier narrowband fluxes are the main cause for erroneous \photoz's, an insight which will inform improvements in forthcoming PAUS data reductions. We have also inspected the spectra and identified catastrophic \specz's, however the effects are shown to be insignificant in this work. Motivated by the study of $30$ outliers shared between \delight and \bcnz, we introduced several new metrics to help improve the identification of \photoz outliers and remove them from the sample to achieve better results. From the $6$ metrics compared, our newly introduced \delight-\bcnz metric ($\Delta_\rf{DB}$) is shown to significantly improve our \photoz quality, allowing it to reach the PAUS target of $\sigma_{68}<0.0035(1+z)$ at $i_\rf{auto}<21.5$ while retaining $60$ per cent of the sample objects. These new metrics could be utilised to return more accurate uncertainties in redshift, which are vital in many cosmological studies.

This opens the door to future studies in finding synergies between different \photoz algorithms and between broadband and narrowband photometry. Together with the promising developments of deep learning approaches to deal with narrowband data \citep{eriksen_pau_2020}, these insights will pave the way towards unprecedentedly precise and accurate photometric redshifts for the full PAUS survey and beyond, like the Javalambre Physics of the Accelerating Universe Astrophysical Survey \citep[J-PAS,][]{benitez_j-pas_2014}.

\section*{Acknowledgements}
The authors wish to thank the referee for the helpful and constructive comments. JYHS would like to thank Boris Leistedt for fruitful discussions and the setup of \delight earlier in this work. JYHS also would like to thank Hwee San Lim and Tiem Leong Yoon for assisting in the setup of equipment in Universiti Sains Malaysia where most of the computational work of this paper was completed. JYHS acknowledges financial support from the MyBrainSc Scholarship by the Ministry of Education, Malaysia, a studentship provided by Ofer Lahav, and the Short Term Research Grant by Universiti Sains Malaysia (304/PFIZIK/6315395). JYHS and BJ acknowledge support by the University College London Cosmoparticle Initiative. MS acknowledges funding from the National Science Centre of Poland (UMO-2016/23/N/ST9/02963) and the Spanish Ministry of Science and Innovation through the Juan de la Cierva Formaci\'on programme (FJC2018-038792-I). H. Hildebrandt acknowledges support by a Heisenberg grant of the Deutsche Forschungsgemeinschaft (Hi 1495/5-1) and an ERC Consolidator Grant (no. 770935). H. Hoekstra acknowledges support from the Netherlands Organisation for Scientific Research (NWO) through grant 639.043.512. IEEC and IFAE are partially funded by the Instituci\'o Centres de Recerca de Catalunya (CERCA) and Beatriu de Pin\'os Programme of Generalitat de Catalunya. Work at Argonne National Lab is supported by UChicago Argonne LLC, Operator of Argonne National Laboratory (Argonne). Argonne, a U.S. Department of Energy Office of Science Laboratory, is operated under contract no. DE-AC02-06CH11357.

This project has received funding from the European Union's Horizon 2020 Research and Innovation Programme under the Marie Sk\l odowska-Curie Actions, through the following projects: Latin American Chinese European Galaxy (LACEGAL) Formation Network (no. 734374), the Enabling Weak Lensing Cosmology (EWC) Programme (no. 776247), and Barcelona Institute of Science and Technology (PROBIST) Postdoctoral Programme (no. 754510). 

PAUS is partially supported by the Ministry of Economy and Competitiveness (MINECO, grants CSD2007-00060, AYA2015-71825, ESP2017-89838, PGC2018-094773, PGC2018-102021, SEV-2016-0588, SEV-2016-0597 and MDM-2015-0509). Funding for PAUS has also been provided by Durham University (ERC StG DEGAS-259586), ETH Zurich, and Leiden University (ERC StG ADULT-279396). 

The PAU data centre is hosted by the Port d'Informaci\'o Cient\'ifica (PIC), maintained through a collaboration of CIEMAT and IFAE, with additional support from Universitat Aut\`onoma de Barcelona and the European Research Development Fund (ERDF).

\section*{Data availability}
The data from PAUS (photometry and \photoz's) is currently not yet publicly available. The data from COSMOS were accessed from the ESO Catalogue Facility (\url{https://www.eso.org/qi/}, while the data from zCOSMOS (spectra and \specz's) were accessed from the zCOSMOS database (\url{http://cesam.lam.fr/zCosmos/}). The derived data generated in this research will be shared on reasonable request to the corresponding author.

\bibliographystyle{mnras}
\bibliography{pau_delight}


\bsp
\label{lastpage}
\end{document}